%% file: main_test.tex
\documentclass[journal]{new-aiaa} %for journal papers
\usepackage[utf8]{inputenc}
\usepackage{amsmath}
\usepackage{graphicx}
\usepackage{longtable,tabularx}
\usepackage{subcaption}
\usepackage{comment}
\usepackage{footnpag}	
\usepackage{cleveref}
\usepackage{bm}
\usepackage[section]{placeins}

\usepackage[numbers]{natbib}

\usepackage{makecell}
\usepackage{adjustbox}
\usepackage{todonotes}
\usepackage{booktabs}
\newcommand{\ra}[1]{\renewcommand{\arraystretch}{#1}}
\usepackage{multirow}
\usepackage{nomencl}
\newcommand\blfootnote[1]{
    \begingroup
    \renewcommand\thefootnote{}\footnote{#1}
    \addtocounter{footnote}{-1}
    \endgroup
}
\makenomenclature

\usepackage{pgfplots}
\pgfplotsset{compat=1.17}
 
\begin{document}
\title{A New Monte-Carlo Model for the Space Environment}

\author{ Daniel Jang \footnote{\textit{Corresponding Author}. Ph.D. Candidate, Department of Aeronautics and Astronautics, Massachusetts Institute of Technology, MA 02139, USA. email: djang@mit.edu},
\ Davide Gusmini \footnote{Visiting Graduate student, Department of Aeronautics and Astronautics, Massachusetts Institute of Technology, MA 02139. email: gusmini.dav@gmail.com} 
\ Peng Mun Siew \footnote{Research Scientist, Department of Aeronautics and Astronautics, Massachusetts Institute of Technology, MA 02139, USA. email: siewpm@mit.edu}}
\affil{Massachusetts Institute of Technology, Cambridge, MA 02139}
\author{Andrea D'Ambrosio \footnote{Postdoctoral Associate, System \& Industrial Engineering Department, University of Arizona, Tucson, AZ 85721, USA; email: dambrosio@arizona.edu}}
\affil{University of Arizona, Tucson, AZ 85721}
\author{Simone Servadio \footnote{Assistant Professor, Department of Aerospace Engineering, Iowa State University, IA 50011, USA. email: servadio@iastate.edu}}
\affil{Iowa State University, IA 50011}
\author{Pablo Machuca\footnote{Visiting Professor, Department of Aerospace Engineering, San Diego State University, CA 92182, USA. email: pmachuca@sdsu.edu}}
\affil{San Diego State University, CA 92182}
\author{Richard Linares \footnote{Rockwell International Career Development Professor, Associate Professor of Aeronautics and Astronautics, Department of Aeronautics and Astronautics, Massachusetts Institute of Technology, Cambridge, MA 02139, USA. email: linaresr@mit.edu}}
\affil{Massachusetts Institute of Technology, Cambridge, MA 02139}

\maketitle
 
 \begin{abstract}

This paper introduces a novel Monte Carlo (MC) method to simulate the evolution of the low-earth orbit environment, enhancing the MIT Orbital Capacity Analysis Tool (MOCAT). 
\blfootnote{ A portion of this work was presented at the 33rd AAS/AIAA Space Flight Mechanics Meeting (Austin, Texas, January 15-19 2023), paper number AIAA-2023-240 }

In recent decades, numerous space environment models have been developed by government agencies and research groups to understand and predict the dynamics of space debris. Our MC approach advances this by simulating the trajectories of space objects and modeling their interactions, such as collisions and explosions. This aids in analyzing the trends of space-object and debris populations. A key innovation of our method is the computational efficiency in orbit propagation, which is crucial for handling potentially large numbers of objects over centuries. We present validation results against the IADC (Inter-Agency Space Debris Coordination Committee) study and explore various scenarios, including ones without future launches and those involving the launch of proposed megaconstellations with over 80,000 active payloads. With the improvement in computational efficiencies provided by this work, we can run these new scenarios that predict millions of trackable objects over a 200-year period. The previous state-of-the-art was 400,000 objects over the same period of time. Notably, while fewer megaconstellations are planned for altitudes above 800 km, even minimal failures in post-mission disposal or collision avoidance maneuvers can significantly impact orbital debris accumulation.

\end{abstract}

\input{nomenclature}

\section{Introduction} 

Space debris has emerged as a critical environmental and political issue due to the substantial increase in space objects, especially in the Low Earth Orbit (LEO).  Several factors contribute to this recent increase in the number of objects, including cost-efficient launches, increased commercial activity, and recent debris-creating events in space, such as explosions and collisions, including anti-satellite weapon tests. In particular, recent advances in orbital launch technologies and the growth of commercial launch providers have made LEO launches much cheaper and more reliable, leading to a new space age. Until recently, the increase in object number in space was around 300 objects per year; however, increased launch cadence and the rise of large LEO constellations have led to a marked increase in LEO population. For example, from 2019 to 2023, SpaceX alone has launched more than 5,500 satellites \cite{SpaceX2022}. The US Federal Communications Commission (FCC) and the United Nations International Telecommunication Union (ITU) filings show that companies or governments are asking for approvals for constellations that are greater than the current number of objects in space. 

As debris from fragmentation events create more debris, it may be possible to create a cascading effect of chain reactions, which is a phenomenon known as Kessler syndrome \cite{Kessler1978, Kessler2010}, which may render certain altitudes unusable and threaten the sustainability of space operations.  The primary natural sink to remove space debris from LEO is the upper atmosphere's low density; there is a limit to how much collision can be tolerated in LEO, which is highly dependent on the orbital altitude. 
% POTENTIAL SOLUTIONS
Recent advances in technologies and policies addressing the space debris problem encompass several key areas. Active Debris Removal (ADR) technologies are being developed, including systems such as harpoons, nets, robotic arms, and space tugs, to capture, deorbit, or mitigate the effects of space debris \cite{ Liou2011, White2014AnRemoval,Hakima2018AssessmentDebris}. 

Addressing the space debris problem requires global collaboration, responsible space practices, effective debris mitigation strategies, and the advancement of technologies for active debris removal and improved space situational awareness (SSA).  In particular, space debris models guide policies and strategies for space traffic management, debris mitigation, and spacecraft design to ensure the sustainability of space activities.  Over the years, studies have shown the dangers of this increased population, and policies and metrics have been proposed to counteract the increased risk in the space environment \cite{FCCdebris2022, SWFhandbookForNewActors2017, SSCbestPractices2019, AIAAbestPractices2022, SPD3_2018, ESAspaceEnvReport2022,colvin2023cost}.  % Computing the LEO orbital capacity requires an understanding of the risk of collisions and the stability of the LEO environment.  
Recently, debris mitigation policies such as the `5-year rule' \cite{FCC5yrRule} and resolutions to limit debris-causing antisatellite tests have come from the United Nations and the commercial space industry \cite{antiASAT2022UN,antiASAT2024SWF}.   

To understand the complex dynamics of space debris, sophisticated modeling approaches have been developed in the literature.  Kessler’s paper in 1978 originally described the potential for runaway growth of orbital debris due to debris that causes more debris through collisions, which could lead to an unusable orbital environment. % \cite{Kessler1978}. 
Since then, several analytical methods have been proposed in the literature to better quantify this risk \cite{Letizia2017,kragCapacity,MITRI}, which can be divided into a few categories.  There are largely two methods to model the evolution of the LEO orbital population and collision risk: statistical sampling methods such as Monte Carlo methods and source-sink models, also known as particle-in-box models.  There are also heuristic metrics to quantify the risk per object for any particular composition of the LEO environment, which is described below.

Space agencies and organizations often use heuristic methods and metrics for decision making when planning missions or strategies to ensure space sustainability.  These indicators are metrics used to assess the potential danger posed by an orbital object and help prioritize the management and mitigation of space debris by evaluating the risk associated with individual debris objects.  
The Criticality of Spacecraft Index (CSI) ranks the environmental criticality of abandoned objects in LEO \cite{Rossi2015}. It takes into account the physical characteristics of a given object, its orbit, and the environment in which it is located. Environmental Consequences of Orbital Breakups (ECOB) is based on the evaluation of the consequences of the fragmentation of the studied object in terms of the increase in the collision probability for operational satellites. This index considers the likelihood and consequence of fragmentation and end-of-life mitigation strategies \cite{Letizia2016severity,Letizia2017}.

In \cite{mcknight2021} numerous international space organizations contributed lists of the top 50 concerning objects, which were compared using multiple algorithms to create a ranked composite list.  Factors such as mass, encounter rates, orbital lifetime, and proximity to operational satellites were shown to be crucial. 

Similarly, the MIT Risk Index (MITRI) is an index to help identify the most dangerous debris that can be removed considering the requirements of the chaser spacecraft and the constraints of the mission \cite{MITRI}.  The index considers the proximity of the debris to highly populated regions, its persistence in orbit, its likelihood to collide, and the estimated number and mass of debris it can generate.

The source-sink evolutionary models describe the interactions between objects populations with ordinary differential equations, first used to describe predator prey interaction between fish population \cite{volterra1926fluctuations} and the chemical reactions \cite{lotka1920analytical}.  For example, if all space objects of interest are categorized as payloads, derelict satellites, or debris, three ordinary differential equations can describe the interaction between these populations.  Average values are often used to describe the population's characteristics, such as a population's size, velocity, rate of launch, and failure rate.  This simplification eliminates the need for computationally expensive propagation of individual object states to estimate a future debris environment.  Gross populations are propagated forward according to the governing differential equations, allowing fast solutions even far into the future.  Exploring a wide set of initial conditions and parameters is much more approachable using such methods.  These methods are also usually deterministic in that, given a set of input variables, a consistent output can be expected as the variables interact in a formalized manner. 

Since Kessler and Cour-Palais first described the feedback runaway phenomenon and identified the risk of an exponential increase in the number of space debris, a few evolutionary models have been proposed in the literature.  Talent introduces the particles-in-box (PIB) model in which a population within an orbital shell is assumed to have some average characteristic and interactions \cite{talent1990pib}.  Fast Debris Evolution (FADE) used simplified first-order differential equations to describe the population interaction \cite{Lewis2009}.  JASON describes a three-population model for one shell and a given launch cadence \cite{JASONreport2020TheSatellites}.  Many models have expanded the evolutionary model to analyze multiple shells, optimal control schemes, and economic equilibrium for maximum policy intake.  The MIT Orbital Capacity Tool Source-Sink Evolutionary Model (MOCAT-SSEM) is able to create a flexible and modular multifidelity model to rapidly model the evolution of the LEO population \cite{MOCATSSEM2023}. The low computational cost demonstrates the ability to optimize user-defined cost functions for policy-making and governance, and calculate the risk-based space environment capacity \cite{d2024carrying} and has been expanded to include orbit-raising dynamics \cite{gusmini2024effects}, as well as calculation of risk-based orbital capacity \cite{d2023novel}.

Statistical sampling methods propagate every object's orbital states with high fidelity propagators to estimate the future space environment at some small time steps, much like a particle filter.  
Several such sampling-based models have been developed by space agencies and private entities as a result of the large-scale development and support required.  Examples include NASA’s Orbital Debris Engineering Model (ORDEM) \cite{Krisko2014TheInvited} and LEO to Geosynchronous Orbit Debris model (LEGEND) \cite{LEGEND2004}, University of Southampton and United Kingdom Space Agency's Debris Analysis and Monitoring Architecture for the Geosynchronous Environment (DAMAGE) \cite{Lewis2001, Lewis2011}, Chinese Academy of Sciences’ SOLEM (Space Objects Long-term Evolution Model) \cite{Wang2019AnSOLEM}, MEDEE model from Centre National d’Etudes Spatiales \cite{medee}, DELTA model from European Space Agency \cite{DELTA2}, LUCA model from Technische University at  Braunschweig \cite{luca2}, NEODEEM model from Kyushu University and the Japan Aerospace Exploration Agency \cite{KAWAMOTO2018}, and others \cite{Drmola2018KesslerModel, Rosengren2019DynamicalOrbits}.  A probabilistic debris environment propagator has been explored in \cite{Giudici2024} where objects are classified as either intact objects or fragments and accounts only for collisions between the two classes.  This simplification allows for computational efficiency, and validation work is on-going.

For each of these models, the debris population and densities are outputted given some input initial conditions and assumptions.  However, computing a debris environment with different sets of assumptions requires high computational cost, as each object must be propagated.  Sampling over a distribution of uncertainties in states and parameters would require an exponential number of propagations.  The high cost is due to the small time steps required to accurately model a collision and semi-analytical propagators requiring high compute cost to propagate far into the future.  For each collision or fragmentation event, a breakup model, such as the NASA Standard Breakup Model (SBM) \cite{nasaSTMevolve4} is used to model the debris cloud generation.  Although the outputted debris distribution for some assumed initial conditions and future traffic models exists, all of these models are closed-source and inputting arbitrary assumptions is difficult if not impossible.

The establishment of a common, validated, and open-source model for space debris is imperative for several reasons. Firstly, such a model ensures standardization and consistency between stakeholders, creating a shared language and methodology for analyzing space debris. This framework facilitates effective communication and information sharing among governments, space agencies, researchers, and industry players. In addition, a validated model improves accuracy in assessing collision risks and predicting debris behavior, crucial for ensuring the safety of spacecraft, satellites, and astronauts in orbit. 

In this paper, an MC method is developed and described to simulate the evolution of the LEO environment called MOCAT-MC.  It is a full-scale three-dimensional debris evolutionary model that propagates individual objects and models the interactions between objects at each time step, with the aim of assessing the LEO population. The development of MOCAT-MC will provide an open-source validated tool that can be accessed and used by the space-debris community.  

One of the drawbacks of MC methods is the high computational time required to run the simulations, and particular attention is given to the computational speed of all sub-modules of MOCAT-MC. Several phenomena are considered, such as the atmospheric model and the propagator, active satellite station-keeping, new launches, reentry, post-mission disposal (PMD), explosions, and collisions.  Active payloads maintain orbital altitude to counteract the effects of atmospheric drag, while new launches are introduced into the simulation. 

Certain payloads are deliberately deorbited through PMD and removed from the simulation due to atmospheric reentry.  A successful PMD is assumed to include the entire deorbit cycle from mission orbit to atmospheric reentry.  However, some satellites may fail to execute PMD with some given probability, transitioning into an inactive state, and remaining in its orbit as a derelict object.

Explosions are simulated with a predefined probability, leading to the creation of smaller debris as described by the NASA Standard Breakup Model. The characteristics of these newly generated objects, such as quantity, direction, and size, are determined by the model.  The Cube method is used to determine collisions between two objects \cite{LiouCUBE2003,LiouCUBE2006}, and results in the generation of numerous debris objects within the simulation as dictated by the NASA SBM.

This work makes several contributions to the literature, including: 
\begin{itemize}
    \item First open-source MC-based evolutionary orbital population model with modular sub-components to allow for numerous propagations, and collision detection algorithms.
    \item Simulation of tens of millions of orbital objects centuries into the future using a single processing thread.    
    \item Demonstration that analytical propagation can be used to scale MC simulations.
    \item Output of collision statistics and collision avoidance burden.
    \item First analysis of future constellation traffic as filed with ITU and FCC, reaching 82000 operational satellites in LEO from the megaconstelations alone.
\end{itemize}

MOCAT was developed in \textsc{Matlab} and is open-source \cite{MOCATMCgithub}. 
The paper is organized as follows.  The MOCAT-MC model is explained in Section \ref{sec:method}.  Section \ref{sec:validation} shows the MOCAT-MC validation exercise against existing scenarios and results in the literature, and Section \ref{sec:results} shows the result from the no future launch scenario, as well as the FCC/ITU filed megaconstellations.  Section \ref{sec:conclusion} summarizes the contribution of this paper.  

% \clearpage

\section{Methodology} \label{sec:method}
\begin{figure}[!ht]
    \makebox[\textwidth][c]{\includegraphics[width=\textwidth]{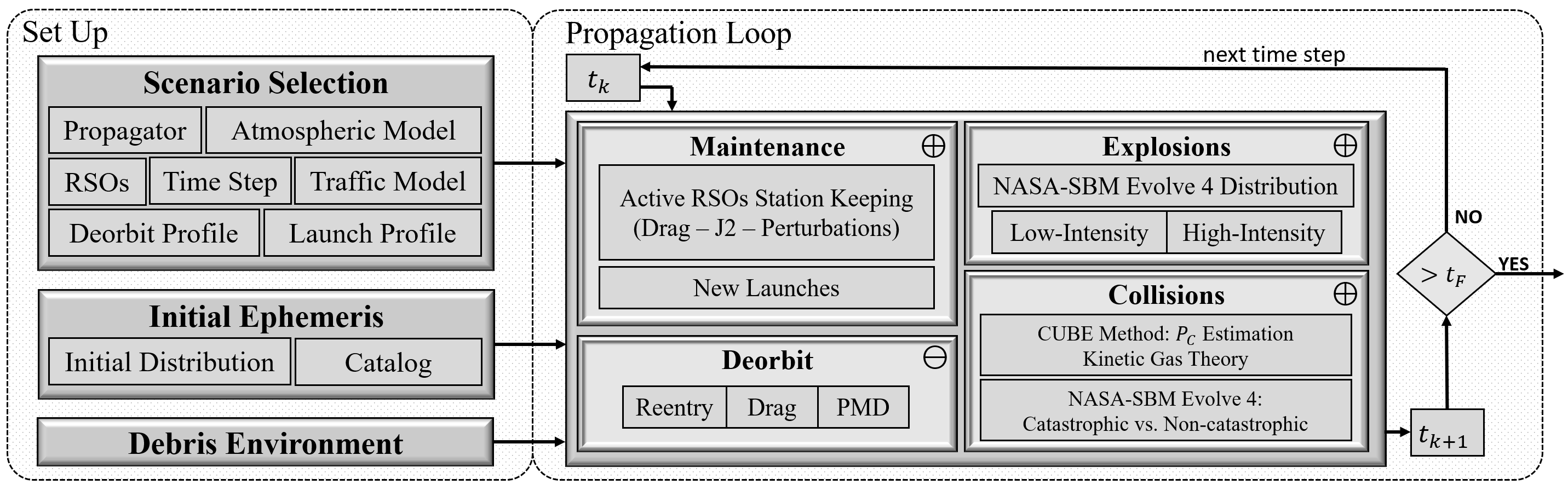}}
    \caption{Schematic of the Monte Carlo tool MOCAT-MC}
    \centering
    \label{fig:schematic-MC}
\end{figure}

MOCAT-MC has multiple submodules, which is shown in the functional diagram in Fig. \ref{fig:schematic-MC} as described in \cite{MITRI}.  The description of each of these components is described in this section.  

\subsection{Analytical Propagator}
The SGP4 propagation method \cite{valladoSGP42006} is a common approach for analytically propagating orbital states.  However, other semi-analytical and analytical propagators could represent a better choice in terms of a compromise between fidelity and computational time. Therefore, in MOCAT-MC, an analytical approximation of the solution for the motion of the LEO objects is used, with perturbations from atmospheric drag and J$_2$. The initial ephemeris is also loaded, which can be seeded with an existing catalog such as the Two Line Element (TLE) catalog provided publicly by the 18th Space Defense Squadron\footnote{\url{https://www.space-track.org}}.

An averaging perturbation technique has been employed to obtain the variational equations for the orbital elements with the combined effect of J$_2$ and drag as described in \cite{Martinusi2015}.  Implementation of other propagators such as SGP4 and the Draper Semi-analytical Satellite Theory (DSST) \cite{cefola2014revisiting} is reserved for future work, though these will require a higher computational burden compared to the analytical propagator.  The propagator is based on two assumptions: the atmospheric density is constant and the orbit eccentricity is small. However, in the implementation of the analytical propagator, a time-varying atmospheric model is modeled with a piecewise continuous formulation with a mean solar activity \cite{Vallado2022} rather than a static atmospheric density. 
This allows for the effects of geomagnetic storms and solar cycles to be modeled, which can strongly affect the atmospheric density.  

Let us define
\begin{align}
    \alpha_0 = \dfrac{\bar{e}_0}{\sqrt{\bar{a}_0}}~, && \beta_0 = \dfrac{\sqrt{3}}{2}~\bar{e}_0, && \bar{n}_0=\sqrt{\frac{\mu}{a_0^3}};
\end{align}
and indicate with $c=\cos{\bar{i}}$ the cosine of the inclination. The resulting set of equations used for the propagation is
\begin{equation}
    \begin{aligned}
      % semi-major axis
      \bar{a} &= \dfrac{\bar{a}_0}{\beta_0^2} \tan^2{\left[ \arctan (\beta_0) - \beta_0 \bar{n}_0 \bar{a}_0 C_0 (t-t_0)\right]}\\
      % eccentricity
      \bar{e} &= \dfrac{2}{\sqrt{3}} \tan{\left[ \arctan (\beta_0) - \beta_0 \bar{n}_0 \bar{a}_0 C_0 (t-t_0)\right]}\\
      % mean anomaly
      \bar{M}-\bar{M}_0 &= \dfrac{1}{8}\dfrac{1}{C_0} \left. \left[\dfrac{4}{\tau} + 3 \alpha_0^2 \ln{\left( \dfrac{\tau}{\bar{a}_0} \right)}\right] \right|_{\tau=\bar{a}_0}^{\tau=\bar{a}} 
      + \dfrac{3k_2 \left(3c^2-1 \right)}{16\mu}\dfrac{1}{C_0} \left. \left[ \dfrac{3 \alpha_0^2}{2} \dfrac{1}{\tau^2} + \dfrac{4}{3 \tau^3}\right] \right|_{\tau=\bar{a}_0}^{\tau=\bar{a}}\\
      % periapsis anomaly
      \bar{\omega}-\bar{\omega}_0 &= \dfrac{3k_2 \left(5c^2-1 \right)}{16\mu} \dfrac{1}{C_0} \left. \left[ \dfrac{5 \alpha_0^2}{2} \dfrac{1}{\tau^2} + \dfrac{4}{3 \tau^3}\right] \right|_{\tau=\bar{a}_0}^{\tau=\bar{a}}\\
      % RAAN
      \bar{\Omega}-\bar{\Omega}_0 &= - \dfrac{3k_2 c}{8\mu} \dfrac{1}{C_0} \left. \left[ \dfrac{5 \alpha_0^2}{2} \dfrac{1}{\tau^2} + \dfrac{4}{3 \tau^3}\right] \right|_{\tau=\bar{a}_0}^{\tau=\bar{a}}\\
    \end{aligned}    
    \label{eq:propagator}
\end{equation}
where $[\bar{a},\bar{e},\bar{i},\bar{\Omega},\bar{\omega},\bar{M}]^T$ represents the state vector at the current time $t$: semi-major axis, eccentricity, inclination, right ascension of the ascending node, argument of periapsis, and mean anomaly, respectively. The subscript $0$ denotes the state variables at the initial time of propagation $t_0$, and $k_2=\mu \textrm{J}_2 R_E^2/2$ 
and $C_0=\frac{1}{2} C_D \frac{A}{m} \rho_0$, with $\mu$ and $R_E$ representing the gravitational parameter and radius of the Earth, respectively, $\rho$ the atmospheric density, and $C_D$, $A$, and $m$ representing the drag coefficient, area, and mass, respectively.

\begin{figure}[!ht]
    \centering
	\subcaptionbox{Analytical propagator\label{fig:still_orbit_an}}[.35\textwidth]{\includegraphics[width=0.9\linewidth]{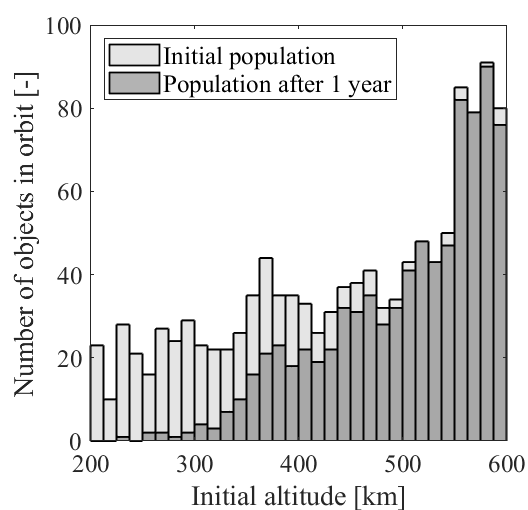}}	
	\subcaptionbox{Numerical propagator\label{fig:still_orbit_dyn}}[.35\textwidth]{\includegraphics[width=0.9\linewidth]{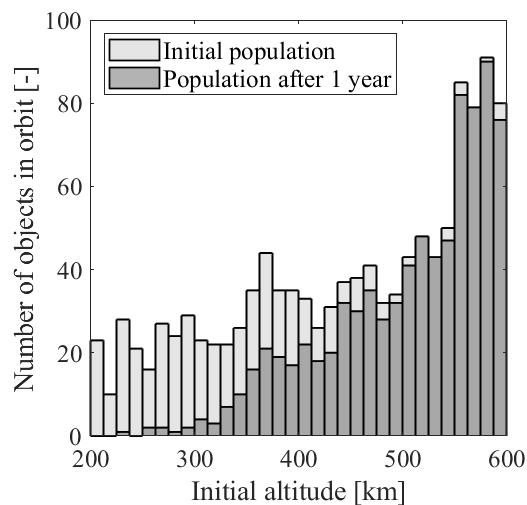}}		  
	\caption{Number of objects in orbit depending on initial altitude}
	\label{fig:still_orbit}
\end{figure}

\begin{figure}[!ht]
    \centering
	\subcaptionbox{Analytical propagator\label{fig:decay_SMA_an}}[.35\textwidth]{\includegraphics[width=0.9\linewidth]{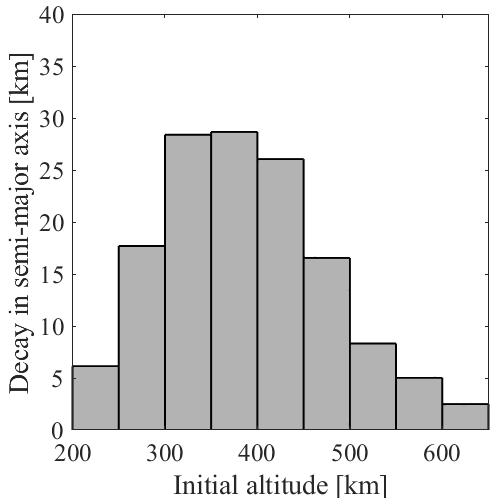}}	
	\subcaptionbox{Numerical propagator\label{fig:decay_SMA_dyn}}[.35\textwidth]{\includegraphics[width=0.9\linewidth]{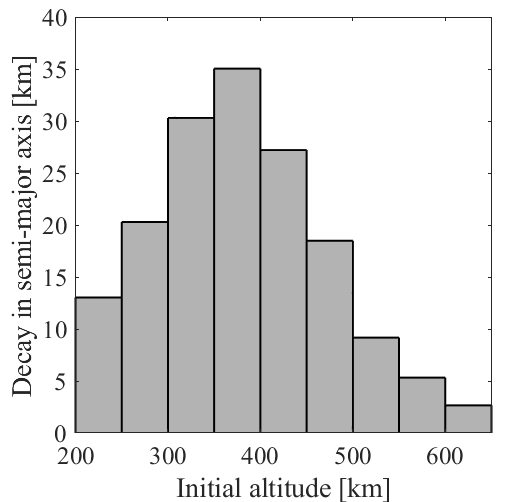}}		  
	\caption{Decay in semi-major axis depending on initial altitude}
	\label{fig:decay_SMA}
\end{figure}

\begin{figure}[!ht]
    \centering
	\subcaptionbox{Analytical propagator\label{fig:decay_RA_an}}[.35\textwidth]{\includegraphics[width=0.9\linewidth]{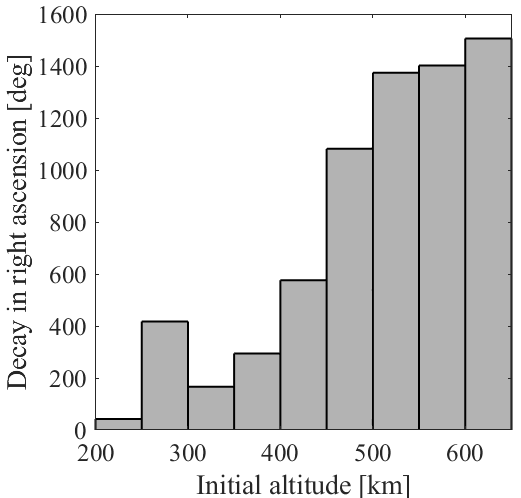}}	
	\subcaptionbox{Numerical propagator\label{fig:decay_RA_dyn}}[.35\textwidth]{\includegraphics[width=0.9\linewidth]{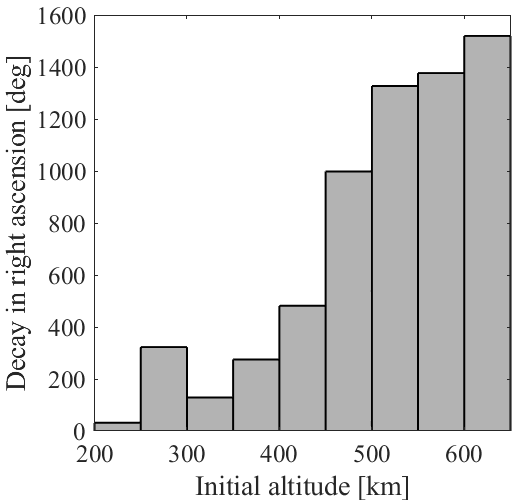}}		  
	\caption{Decay in right ascension depending on initial altitude}
	\label{fig:decay_RA}
\end{figure}

\begin{figure}[!ht]
    \centering
	\subcaptionbox{Analytical propagator\label{fig:time_decay_an}}[.35\textwidth]{\includegraphics[width=0.9\linewidth]{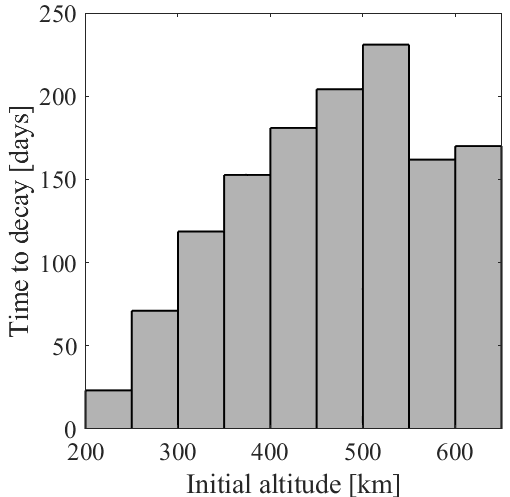}}	
	\subcaptionbox{Numerical propagator\label{fig:time_decay_dyn}}[.35\textwidth]{\includegraphics[width=0.9\linewidth]{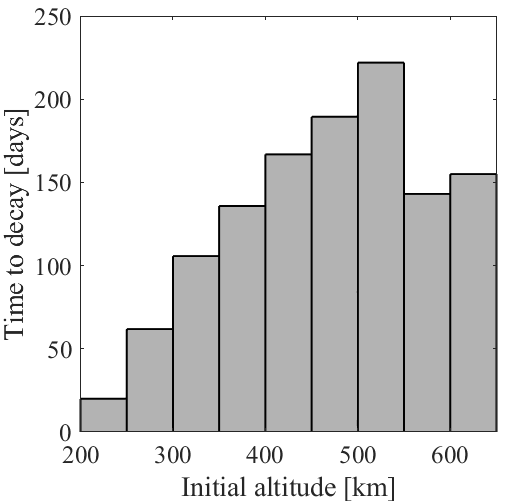}}		  
	\caption{Time to decay depending on initial altitude}
	\label{fig:time_decay}
\end{figure}

Before introducing the time-varying model of atmospheric density in the next section, a static exponential atmospheric model is used to test the validity of the analytical equations of motion in Eq. \eqref{eq:propagator} compared to a numerical propagator that propagates orbits with direct numerical integration of the equations of motion including the drag and J$_2$ perturbations. For a given initial population of objects between 200 and 2000 km altitudes and a propagation duration of one year, several relevant metrics are used for validation purposes: number of objects still in orbit depending on initial altitude $h_0$; reduction in the semi-major axis depending on $h_0$ and decay in right ascension depending on $h_0$; and finally the time to decay depending on $h_0$.  Figures \labelcref{fig:still_orbit,fig:decay_SMA,fig:decay_RA,fig:time_decay} show the validation results.  

Figure \ref{fig:still_orbit}, for instance, illustrates the number of objects as a function of altitude, in the initial population and after the one-year propagation. It can be observed that the distributions of objects obtained by the analytical propagator and by the numerical propagator show strong agreement.  The analytical propagator can effectively approximate the number of objects expected as a function of altitude. From Fig. \ref{fig:decay_SMA}, it is shown that the analytical propagator typically underestimates the decay in the semi-major axis due to drag, but the overall distribution of semi-major decay as a function of altitude resembles that of the numerical propagator. On the contrary, in Fig. \ref{fig:decay_RA}, it is observed that the decrease in right ascension is usually overestimated by the analytical propagator, but the overall distribution of the decrease in right ascension as a function of altitude is also similar to that displayed by the numerical propagator. Lastly, and related to the results in Fig. \ref{fig:decay_SMA}, it is observed in Fig. \ref{fig:time_decay} that the time to decay is usually overestimated by the analytical propagator, but the distribution of time to decay as a function of altitude produced by the analytical propagation also resembles that produced by the numerical propagator.  Although certain quantitative discrepancies appear between the analytical and numerical propagators, the analytical solution is able to capture the overall effects of drag and J$_2$ as a function of time, while achieving orders of magnitude shorter computational times: which is particularly meaningful when propagating tens of thousands to millions of orbital objects centuries into the future.  This is also in-line with \cite{Martinusi2015} where the largest component of error was due to the averaging of the J$_2$ effect, which is less of a concern for long-term collision models such as MOCAT-MC.  

\subsection{Initial Population and Data Sources}

Simulations that have a basis in today's orbital environment require some data to seed the initial population.  A range of orbital parameters can be provided to the simulation, which can be sampled to seed the initial orbital distribution.  Each object can have a unique lifetime, station-keeping methods, failure rate, size, etc. to characterize its behavior and potential interaction with other objects.  The initial debris environment is determined.  Depending on the analysis, the size, number, and orbital parameters of these objects can be set. The minimum size debris to be considered in the model will also be an input parameter, which will affect the number of objects in the simulation.  The probability distribution function (PDF) of the debris population parameters can be supplied from LEO debris models such as ESA's MASTER model \cite{MASTER8} and the current space catalog.  

Two Line Element sets (TLEs) are used to identify objects that are currently in orbit around Earth. The model considers a total of around 24000 space objects obtained from Space-Track (as of September 2022)\footnote{\url{https://www.space-track.org}}. 
In order to have a complete dataset, some other information such as mass, diameter, status, object-class, and launch date need to be gathered.  The physical characteristics, the date of launch, and the object-class are retrieved through DISCOS\footnote{\url{https://discosweb.esoc.esa.int}} developed by the European Space Agency.  For up-to-date active/inactive status of payloads CelesTrak  \footnote{\url{https://celestrak.org}} 
is used.  This may be important, as active payloads are able to perform station keeping and collision avoidance maneuvers and may attempt to dispose after their mission is over.  After processing the data, MIT's catalog with an epoch of January 1 2022 consists of 21014 TLEs, out of which 10869 are debris, 7015 payloads of which 5129 active and 1886 inactive, 1421 rocket bodies, and the remaining 1709 Mission Related Objects (MRO).

\begin{table}[htp!]
\centering
\caption{Definition of parameters for each orbital object in MOCAT-MC}
\begin{tabular}{@{}lccccccc@{}}
\toprule
Column & 1 - 6        & 7          & 8          & 9     & 10     & 11     & 12       \\ \midrule
Description & $\bar{a},\bar{e},\bar{i},\bar{\Omega},\bar{\omega},\bar{M}$       & $B^*$     & Mass           & Radius       & \begin{tabular}[c]{@{}c@{}}Error \\ flag\end{tabular} & \begin{tabular}[c]{@{}c@{}}Control \\ flag\end{tabular} & $a_{desired}$ \\ \bottomrule  \toprule
Column & 13    & 14              & 15 - 16           & 17 - 19        & 20 - 22           & 23             & 24          \\ \midrule
Description & \begin{tabular}[c]{@{}c@{}}Mission \\ duration\end{tabular} & \begin{tabular}[c]{@{}c@{}}Constel \\ number\end{tabular} & \begin{tabular}[c]{@{}c@{}}Date created, \\ date launched\end{tabular} & $r_{x,y,z}$ & $v_{x,y,z}$       & \begin{tabular}[c]{@{}c@{}}Object \\ class\end{tabular} & ID          \\ \bottomrule
\end{tabular}
\label{tab:getidx}
\end{table}

The parameters defined and tracked for each object throughout the simulation are defined in Table \ref{tab:getidx}.  The mean orbital elements are defined with semimajor-axis, eccentricity, inclination, RAAN, argument of perigee, and mean anomaly as:
$[\bar{a},\bar{e},\bar{i},\bar{\Omega},\bar{\omega},\bar{M}]$.  $B^*$ is an adjusted value of the ballistic coefficient of the satellite.  The mass is denoted in kg, and radius is denoted in meters.  The error flag is used for the internal propagation error state.  The control flag denotes whether the object has control to stay at the desired semi-major axis $a_{desired}$ for the mission duration after launch.  The constellation number denotes which constellation the object is part of, as defined in the constellation input file.  The date created and the date launched are defined in modified Julian day (MJD). $r_{x,y,z}, v_{x,y,z}$ are cartesian states in inertial frame (ECI).  The object class denotes the type of object, as described in Table \ref{tab:objectclass-def}.  These definitions are consistent with those of the ESA MASTER database \cite{MASTER8}. The satellite ID is defined for the purpose of tracking individual objects throughout the simulation.

\begin{table}[!htp]
    \centering
    \caption{Definition of object-class in MOCAT-MC}
    \begin{tabular}{@{}lccccccccccc@{}}
        \toprule
        Class & 1 & 2 & 3 & 4 & 5 & 6 & 7 & 8 & 9 & 10 & 11\\ \midrule
        Definition & PL & PL MRO & PL FD & PL D & RB & RB MRO & RB FD & RB D & D & Other D & Ukn\\ \bottomrule
    \end{tabular}
    \label{tab:objectclass-def}
\end{table}

Although all TLE objects are represented in the DISCOS database, some entries omit certain data.  For MOCAT-MC, the DISCOS dataset is used for the physical parameters, launch date (or creation date for debris), and the type of object as one of the 11 categories defined by ESA.  As the TLEs provide orbital parameters and type of object divided into Payload, Rocket Body, Debris, and Unknown, the objects with missing DISCOS data are sampled randomly from the PDF produced by the data from existing equivalent object type using a 2-D Gaussian fit.  ESA's DELTA model resamples the size and mass by assuming an aluminum sphere for the density to calculate the mass from the radius and vice versa, which often overestimates the density of fragmentation debris.  

\subsection{Launch Rate}
The launch profile of new objects can be defined arbitrarily.  Historical launch rates and object parameters can be used, such as the past $n$ years of launches, where those objects are launched into the same orbits with the same object parameters, such as mass and shape.  Some other studies have divided objects into constellation objects, which have a constant launch rate for replenishment level launches assuming some constant desired constellation size, while non-constellation objects would be launched at the historical rate over the $n$ year period.  Another method of launching would be completely arbitrary, where payloads with some lifetime are launched into the orbit of interest with randomized orbital parameters.  Debris and rocket bodies can also be introduced per launch, as desired.  For both of these scenarios, an assumed increase in launch rate per year can be specified, modeling some linear increase in launch activity with respect to time.  The FCC and ITU filings for future megaconstellation have been compiled and are discussed in Sec. \ref{sec:Megaconstellations}.

\subsection{Atmospheric Density Modeling} 

LEO objects are strongly perturbed by the Earth's upper atmosphere due to drag. Hence, accurate propagation of LEO objects requires an accurate estimate of the drag force caused by the Earth's upper atmosphere. However, the ionosphere-thermosphere system is highly dynamic and is strongly influenced by solar and geomagnetic activities. 

In lieu of a static exponential density model, in this work the Jacchia-Bowman 2008 (JB2008) density model is used \cite{bowman2008new}. The JB2008 density model is an empirical density model that is extrapolated based on past historical atmospheric density data to capture the statistically average behavior of the atmosphere under different solar and geomagnetic forcing. The main drivers for the JB2008 density model are a set of solar fluxes measured at different wavelengths and the temperature change due to the Disturbance Storm Time index (DSTDTC). The set of solar fluxes captures the effects of solar activities; the DSTDTC index, on the other hand, captures the effect of geomagnetic activities on the thermospheric density field.

The difficulties in accurately predicting the long-term solar and geomagnetic indices are one of the main challenges with using the JB2008 density model for long-term prediction. The solar fluxes closely correlate to the solar cycle. A solar cycle lasts approximately 11 years and corresponds to the Sun's magnetic field cycle. 
Here, a moderately active solar cycle is assumed for all future solar cycles, using the observed solar indices for a moderately active solar cycle taken from \cite{tobiska2008solar}. At each propagation instance, the solar indices are sampled as independent Gaussian variables around the mean observed values, depending on the relative month in the solar cycle. On the other hand, a fixed DSTDTC value of 58k is used, which corresponds to the historical long-term mean DSTDTC value. However, note that the strength of the solar cycle tends to vary across different solar cycles and cycles over periods of high activities and low activities, and this is not reflected in our current assumption of consistently moderately active solar cycles.

Note that controlled payloads are propagated such that the semi-major axis will stay relatively constant.  When $a$ of such an object deviates by some defined amount from $a_\text{desired}$, $a$ is simply updated to become $a_\text{desired}$ to simulate station-keeping at the desired altitude.  

\begin{figure}[!htb]
    \centering
    \includegraphics[width=0.6\textwidth]{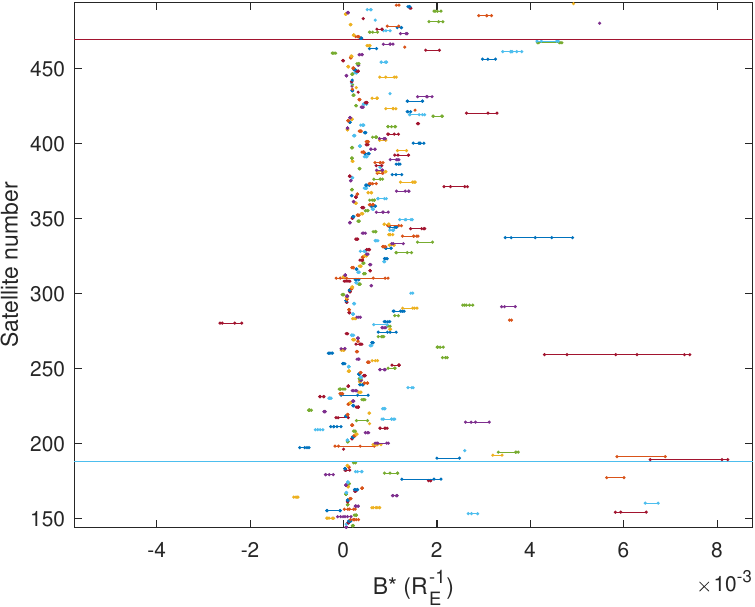} 
    \caption{Variability of $B^*$ values from Space-track.org TLE's across a 5 day span}
    \label{fig:BstarDistribution1}
\end{figure}

The $B^*$ value can be defined or calculated in several ways.  The $B^*$ value is provided in a field in the TLE catalog.  Note that the provided value may be negative and nonphysical, as this parameter is often a free parameter that is fitted to the sensor data to fit an orbit to create the TLEs.  Figure \ref{fig:BstarDistribution1} shows the distribution of some TLE sampling over a period of several days, showing the prevalence of negative $B^*$ values and the magnitude with which the value varies from epoch to epoch.  Negative $B^*$ values are non-physical, and such variability comes from the fact that it is an arbitrary free parameter in differential correction schemes used in producing the TLEs \cite{Vallado2022}. The horizontal lines in Fig. \ref{fig:BstarDistribution1} denote the values of $B^*$ that span more than the limits of the axis, while the markers show the values from the multiple TLEs in that time span.  This shows a wide range of the $B^*$ values even for the same object within a week as reported by the TLE.  A histogram of the $B^*$ values from TLEs from January 2023 is shown in the Appendix. For non-SGP4 propagators, $B^*$ or AMR should be calculated separately.  MOCAT-MC calculates the AMR from the physical properties of the objects, as noted in the ESA DISCOS database \cite{DISCOS}.  

Alternatively, $B^*$  can be recalculated from the physical characteristics of the satellite.  This can be calculated as:
\begin{equation}
    B^* = \frac{\pi r^2 \cdot C_D\cdot  \rho_0}{2m} 
\end{equation}
where $C_D$ is the ballistic coefficient and is simply defined as $C_D = 2.2$,  $r$ is the radius of the object, $\rho_0$ is the reference air density which is $0.15696615{\text{ kg}}/(\mathrm {m} ^{2}\cdot R_{\text{Earth}})$ \cite{Hoots80}, and the unit for $B^*$ is $R_E^{-1}$. 

\begin{figure}[!h]
    \centering
	\subcaptionbox{Static exponential model}[.35\textwidth]{\includegraphics[scale=0.75]{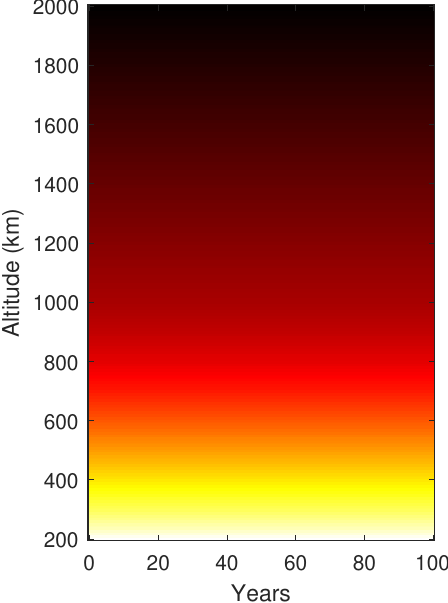}}
	\subcaptionbox{Extrapolated JB2008 model}[.5\textwidth]{\includegraphics[scale=0.75]{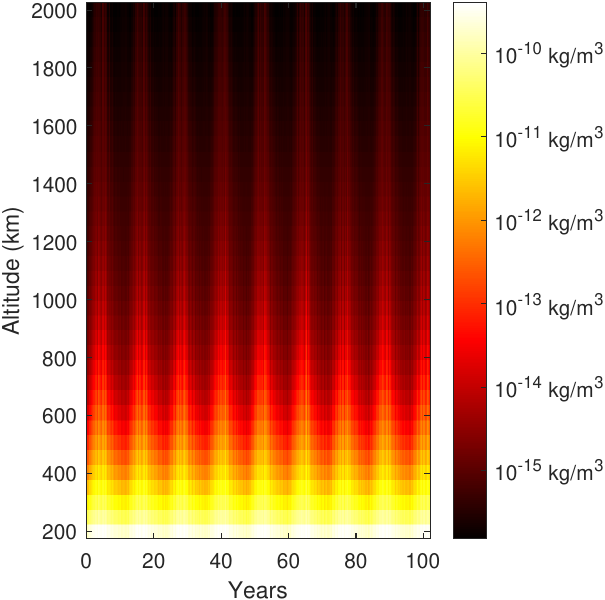}}  
	\caption{Comparison between the two atmospheric models}
	\label{fig:atmosphericModels}
    % atmosphericModels.m
\end{figure}

To include the effects of space weather, in terms of solar and geomagnetic activities, on atmospheric density, the static exponential density model can be replaced by the static density model in \cite{acedo2017kinematics}.  Since the model mentioned above is valid within the range 150-1100 km, the reference altitude is assumed to be fixed at 150 km, for altitudes below 150 km, and at 1100 km, for altitudes above 1100 km.  The comparison between these density models is seen in Fig. \ref{fig:atmosphericModels}.  

\subsection{Collision detection}

There is a spectrum of methods to determine if a collision has occurred, ranging from deterministic methods to stochastic sampling-based methods.  For the stochastic method, the Cube method has been used in numerous MC tools due to its scalability and simple implementation \cite{LiouCUBE2003, LEGEND2004}, although some care is needed to adjust the algorithm parameters to allow realistic collision probability \cite{LewisCUBElimitations2019}.  The computational scale of Cube to $n$ objects is $\mathcal{O}(n)$.  

Deterministic methods calculate the actual point of closest approach for each pair of satellites that are nearby.  Adaptive propagation timesteps are needed to calculate the precise moment and distance between two objects, which imposes a significantly higher computational cost, and generally the computational scaling of these PCA methods to $n$ objects would be $\mathcal{O}(n^2)$; however, methods that filter pairs of comparisons have been proposed in the literature that allow for complexity of $\mathcal{O}(n^{1.5})$ \cite{Rodriguez2002, George2011, Lue2011}.  A comparison between these methods is shown in the results section. Once a collision is detected, fragmentation dynamics using the NASA Standard Breakup Model (EVOLVE4) is used as described in the next section.

\begin{figure}[!htb]
    \centering
	\subcaptionbox{Divide the space into cubes at each timestep}[.45\textwidth]{\includegraphics[scale=0.7]{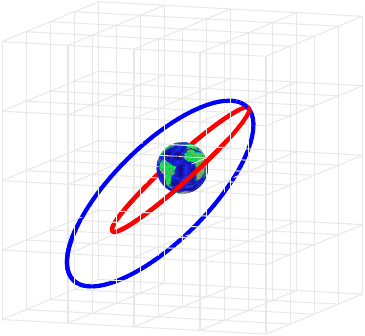}}
	\subcaptionbox{If two or more objects occupy the same cube, approximate the objects as gas particles in the cube volume}[.45\textwidth]{\includegraphics[scale=0.52]{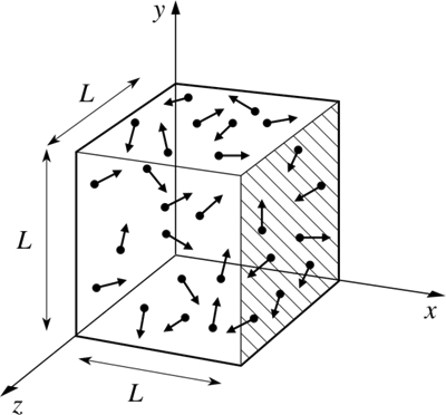}}  
	\caption{The two-step Cube collision detection scheme}
	\label{fig:cubewire}
\end{figure}

The Cube method estimates the long-term collision probability by uniformly sampling the objects over time \cite{LiouCUBE2003}, which is explained here.  The method treats any pair of objects to have a static probability of collision over a long period of time, and the collision probability is calculated for a particular moment in time when the objects are near each other.  This is determined by discretizing the orbital space into cubes and flagging the two objects as close when they are in the same cube.  At this point, the kinetic theory of gas is applied to determine the probability of collision.  The probability of collision is calculated for any two objects that reside within a discretized cube concurrently, as 
\begin{equation}
    P_{i,j} = s_i s_j V \sigma dU
    \label{eq:cubeprobability}
\end{equation}
where $s_i$ and $s_j$ are the spatial densities of objects $i$ and $j$ in the cube, respectively, $V$ is the relative velocity between the two objects, $\sigma$ is the collision cross-sectional area, and $dU$ is the volume of the cube.  Note that this probability is the probability of collision per time, and the aggregate collision for the time step $\Delta t$ is calculated at that time step as $P_{i,j}\Delta t $.  The length of a cube for the proximity filter is often taken as 1\% or less of the mean semi-major axis of objects, which is around 70 km for a population of LEO objects.  The algorithm is described graphically in Fig. \ref{fig:cubewire}, where the two-step process uses a proximity filter before employing the gas particle collision model.  The sensitivity and validity of using the Cube algorithm with various discretized time steps and cube size for collision detection modeling has previously been explored in the literature \cite{LewisCUBElimitations2019}.  The relationship between cube resolution and the estimation of $P_{i,j}$ is explored later in the paper for the purpose of informing MOCAT-MC and its collision probability parameters.  One benefit of the Cube approach is that the computational complexity is $\mathcal{O}(n)$ as opposed to the exhaustive pairwise comparison that yields $\mathcal{O}(n^2)$ for $n$ objects.

\subsection{Fragmentation Model}

Fragmentation events in MOCAT-MC are simulated with the NASA Standard Breakup Model (EVOLVE4).  The NASA SBM is a semi-empirical model based on evidence compiled from historical orbital data measured \textit{in-situ} on-orbit as well as from terrestrial radar measurements and terrestrial hyper-velocity impact experiments.  The model is deterministic and sample-based, and the samples are described by $L_C$ the characteristic length, $A/m$ the area-to-mass ratio, and $\Delta v$ the ejection velocity in a random direction from the parent velocity. 
 
The SBM specifies that the impact energy per target mass is
\begin{equation}
    \Tilde{E}_p = \frac{1}{2}\frac{m_c}{m_t}v_c^2,
\end{equation}
where $m_c$ is the mass of the chaser and $m_t$ is the mass of the target, and impact velocity is $v_c$.  The mass of the target is assumed to be greater than the mass of the chaser.  In the SBM, a collision is considered \textit{catastrophic} where the chaser and the target are completely fragmented when $\Tilde{E}_p > 40$ J/g and \textit{non-catastrophic} if not.  

Ref. \cite{Frey2021} reformulates the model into a PDF, where the number of objects produced with $L$ that is greater than some lower bound $L_0$ is
\begin{equation}
    N_L(L_0) = k L_0^{-\beta} \hspace{2em} k,\beta > 0 \label{eq:FreyPDF}.
\end{equation}
$k$ and $\beta$ are unitless parameters that depend on the type of fragmentation and the physical characteristics of the objects involved.  Although the SBM does not inherently conserve physical quantities such as mass and kinetic energy \cite{Finkleman2008AnalysisEvents}, this formulation of the breakup model into a PDF allows for the conservation of mass and energy to be enforced.  

The model takes as input the smallest characteristic length $L_C$ for the generation of debris objects.  The characteristic length is defined as the mean of the three maximum orthogonal projected dimensions of the object as $L_C = (L_x+L_y+L_z) /3$. 

This is an input variable that can be adjusted; many simulations use the value 0.1 m as this is the commonly-used threshold for trackable objects in LEO. Note that $L_C$ will be treated as an equivalent diameter $d$. Collisions between objects result in the generation of new objects within the simulation. The parameters of these new objects also follow the NASA SBM, as described in Eqs. \ref{eq:NASAsbm} and \ref{eq:NASAsbm2}.  The number of fragments $n_f$ of diameter $d>L_C$ can be computed as 
\begin{equation}
    n_f = \begin{cases}
    6\, c_s \, \hat{L}_C^{-1.6} &\quad \textrm{for explosions}\\
    0.1\, \hat{m}^{0.75} \, \hat{L}_C^{-1.71} &\quad \textrm{for collisions}
    \end{cases}
    \label{eq:NASAsbm}
 \end{equation}

where % $\hat{L}_C = {L}_C / [\textrm{m}]$ and 

\begin{align}
    \hat{m} = \begin{cases}
    \dfrac{m_t+m_p}{\textrm{[kg]}} &\quad \textrm{for } \Tilde{E}_p\geq\Tilde{E}_p^*\\
    \dfrac{m_p\,v_i^2}{\textrm{1000[kg\,(m/s)$^2$]}} &\quad \textrm{for } \Tilde{E}_p<\Tilde{E}_p^*
    \end{cases} && \quad \textrm{with} &&
    \Tilde{E}_p = \frac{m_p\,v_i^2}{2\,m_t}
    \label{eq:NASAsbm2}
\end{align}

Note that the symbol $\,\hat{}\,$ indicates normalized quantities, $m_t$ and $m_p$ are, respectively, the target (mostly derelicts or rocket bodies) and the projectile mass, $v_i$ the relative velocity, $\Tilde{E}_p$ the specific energy of the projectile, and $\Tilde{E}_p^*=40$ [kJ/kg] the specific energy threshold for a catastrophic collision. 

Particular attention should be paid to the scaling parameter $c_s$ in Eq.\:\ref{eq:NASAsbm}. It is an event-specific calibration constant based on historic events, and an empirical correction for certain classes of fragmentation events, with $0.1 \leq c_s \leq 1.0$. For a mass between 600 kg and 1000 kg, the calibration factor is $c_s = 1.0$. However, past fragmentation events showed very different characteristics, thus the break-up models need to be calibrated.  The implementation flow-chart of the NASA SBM in MOCAT-MC is shown in the Appendix. 

\subsection{Area to Mass Ratio Calculation} \label{sec:MCamr}

According to the NASA SBM, the area-to-mass ratio $A/m$ for new fragments is assigned according to a bimodal probability density function $p(\chi, \vartheta)$.
\begin{equation}
    p(\chi, \vartheta) = \alpha(\vartheta)~p_1(\chi) + (1-\alpha(\vartheta))~p_2(\chi)
\end{equation}
where $\chi = \log_{10}(\{A/m \}/[\textrm{m}^2/\textrm{kg}])$ is the area-to-mass parameter, $\vartheta=log_{10}(d/[\textrm{m}])$, and $p_{1,2}$ indicates the normally distributed density functions. The parameter $\alpha$, the means $\mu_{1,2}$, and the standard deviations $\sigma_{1,2}$ are calculated as stated in the NASA SBM.  The effective cross-section $A$, function of the fragment diameter d is
\begin{equation}
    A/[\textrm{m}^2]= \begin{cases}
    0.540424 (d/[\rm m])^2 & \quad \textrm{for}~d<1.67 \textrm{mm}\\
    0.556945 (d/[\rm m])^{2.0047077} & \quad \textrm{for}~d\geq1.67 \textrm{mm}\\
    \end{cases}
\end{equation}
The fragment mass is thus determined as
\begin{equation}
    m = \dfrac{A}{A/m}
\end{equation}
The model requires also to assign the imparted fragmentation velocities, which are sampled from a normal distribution characterised by the following mean value and standard deviation according to the SBM:
\begin{equation}
  \begin{aligned}
    \mu_\nu = 0.2 \chi + 1.85 && \sigma_\nu=0.4 &&& \textrm{for explosions}\\
    \mu_\nu = 0.9 \chi + 2.90 && \sigma_\nu=0.4 &&& \textrm{for collisions}
  \end{aligned}
\end{equation}
where $\nu = \log_{10}(\Delta v/[\textrm{m}/\textrm{s}])$.

The propagation of debris clouds has also been an active area of research.  Nominally, a covariance propagation model may be adapted, though the log-normal distribution of the number of objects and imparted $\Delta V$ in a collision is ill-suited to the multivariate Gaussian assumption for covariances.  The evolution of debris clouds and their effect on key LEO orbits have been formulated and analyzed in \cite{Heard1976DispersionParticles, Cordelli1991TheModel, Frazzoli1996DebrisDesign}.  Efficient cloud propagation has been explored through the DAMAGE model and other MC approaches \cite{Rossi2016AnalysisOrbits, LewisSensitivity, Letizia2018ExtensionPropagation}.  The specific contribution of explosion and future collision fragments to the orbital debris environment showed the importance of mitigating fragmentation events \cite{Su1985ContributionEnvironment, Ruch2021DECOUPLEDENVIRONMENT}.  
The probabilistic uncertainty of the lifetime of orbital debris was analyzed in \cite{DellElce2015ProbabilisticCharacterization, Luo2017AMechanics, LewisUnderstandingDynamics}.  The effect of thermospheric contraction on debris reentry has been modeled \cite{Lewis2011}.  Empirical modeling of fragmentation events such as the Iridium-Cosmos collision and the 2013 ASAT test has validated some models \cite{Weeden2013Anti-satelliteChina, Kelso2009AnalysisCollision, Stansbery2008}. 

\begin{figure}[!ht]
    \centering
	\includegraphics[width=0.5\linewidth]{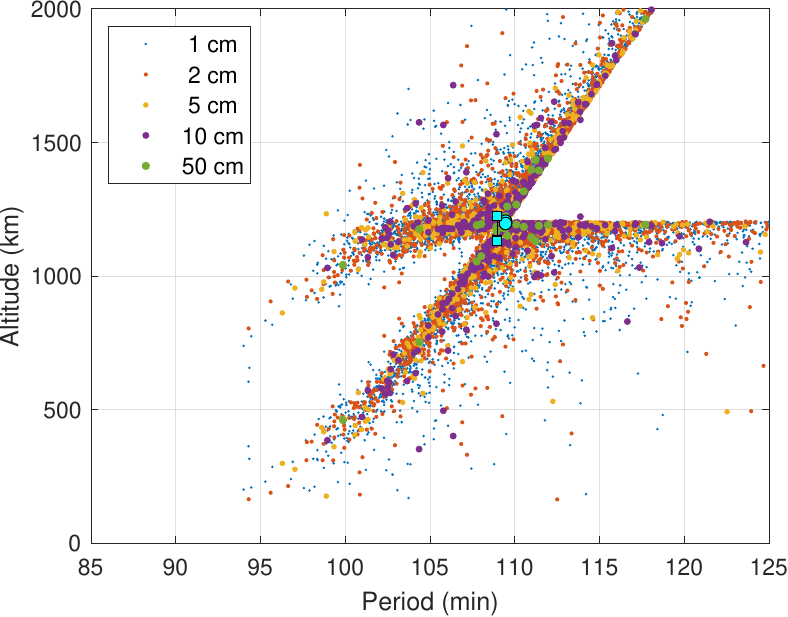}
	\caption{Gabbard plot of debris distribution after a collision}
	\label{fig:frag_binning_MCvsSSEM}
    % djangSBX/frag_test_sweepPs_vec2.m
\end{figure}

The generation and propagation of the debris using MOCAT-MC for a collision is shown as a Gabbard plot in Fig. \ref{fig:frag_binning_MCvsSSEM}, with debris fragments divided into a range of sizes. The collision is between a 200 kg parent object and a 1 kg impactor, and the fragments are generated using the NASA Standard Breakup Model with $L_C = 10$ cm.  The epoch shown is 30 years after the collision.  The propagated locations of the original two objects are shown as teal circles.  As implemented in the NASA SBM, the magnitude of the $\Delta V$ imparted to each of the debris is stochastically assigned as a magnitude.  For momentum conservation, it is assumed that the direction of the magnitude is uniformly distributed spherically, as is also implemented in the literature \cite{BraunAnalysisBreakup2017, Frey2021,servadio2024threat}.

\subsection{Active Payload Orbit Control}

During propagation, each object is subjected to the effects of atmospheric drag, which causes the reduction of the semi-major axis $a$. On the one hand, this consequence is beneficial for LEO safety because it removes derelicts and debris, but on the other hand, it has to be counteracted by active satellites. Depending on the mission altitude and the solar activity level, perturbations, mostly drag effects, could require a satellite to perform station-keeping maneuvers even once per orbit. For current and future satellites, in particular those composing megaconstellations, electric propulsion seems to be the adopted hardware solution. This kind of propulsion system can provide a small level of thrust for a limited amount of time due to limitations in the available electric power. All of these factors, hardware solution, mission design, and mission constraints, could yield a satellite to split a maneuver into a small set of successive submaneuvers over each or a few orbits. %\cite{maisonobe2022very} 
Since orbit control modeling is not the main focus of the paper, the model considers the active satellite semi-major axis variation to be null. In fact, at each time step, all Keplerian orbital elements, except $a$, are propagated forward in time for active payloads. More accurate models are planned to be included in future works.

During propagation, each satellite is subjected to the effects of atmospheric drag. To model station-keeping maneuvers for active satellites, a threshold on the semi-major axis variation $\Delta a_\text{thr}$ has been established. The simplified maneuver consists of adjusting the current semi-major axis $a_\text{current}(t)$ (time dependent) to the initial and desired value $a_\text{desired}$ whenever, during propagation, the following condition is verified.

\begin{equation}
    a_\text{desired}-a_\text{current}(t) > \Delta a_\text{thr}
\end{equation}

Moreover, active satellites are supposed to perform collision avoidance maneuvers with a certain probability of failure, indicated by $\alpha$ in case the satellite encounters a noncooperative species (rocket bodies, derelict or debris), while with $\alpha_a$ if the encounter is between two active satellites. The resulting probability of collision, computed with Eq.\;\ref{eq:cubeprobability} is therefore pre-multiplied by one of these two factors as:
\begin{equation} 
    P_{i,j} = \begin{cases}
    \alpha_a ~ P_{i,j} & \quad \textrm{active-active encounter}\\
    \alpha ~ P_{i,j}   & \quad \textrm{otherwise}
    \end{cases}
\end{equation}

Although these terms are defined by the user, realistically $\alpha_a<\alpha$, as a pair of satellites is assumed to have a greater chance of successfully performing a collision avoidance maneuver through coordination compared to an active satellite against debris.  

Despite the large number of potential conjunction events, there has never been an active-on-active collision, leading to a historical trend of $\alpha_a = 0$. Even an active-on-derelict conjunction is quite rare with such safeguards and warnings in place. The first collision between satellites was the Iridium-Cosmos collision in 2009, although the policy of issuing CDMs using high-quality data had not been implemented then \cite{SWFiridiumcosmos}. 

Although there has been a rapid growth in controlled payloads launched into LEO, collisions between an active satellite have not happened yet.  However, near-conjunctions for active satellites are a daily occurrence. Since the first Starlink satellites were launched in 2019, they have had to perform 50,000 collision avoidance maneuvers by 2023, and at the current rate, the Starlink satellites will have to maneuver more than a million times in a half-year span by 2028 \cite{starlinkManeuver}.  The 18th and 19th Space Defense Squadrons have issued more than 170 million conjunction data messages (CDM) for more than 3 million conjunction events between January 2016 and December 2021 \cite{MOOMEY2023217}.  With an increase in the number of constellation operators, so does the need for proper coordination between the operators.  The increased complexity may result in the first ever collision between two active payloads.

\subsection{Future Constellation Traffic}

Future constellation traffic is described by specifying a few parameters per constellation in an external file that is ingested by the simulation during the setup phase.  Several parameters must be entered for each constellation.  The altitude and inclination are given for each shell or generation of the constellation and the number of satellites already in orbit at the start of the epoch. Satellite parameters such as mass, radius, and mission life are also required, although missing information will default to using Starlink Gen 1 specifications.  The date of start and end of the build-up phase must be supplied to calculate the launch rate during the build-up phase, where a linear build-up phase is assumed.  Throughout the operational phase of the constellation, the appropriate replenishment satellites are launched for the satellite that has reached the end of life.  The end of operation date can be supplied to the constellation, after which no more replenishment satellites are launched.  The constellation company index can be supplied to group the separate shells and phases into the same constellation.  Intra-constellation collision avoidance efficacy, for example, can be specified such that the collision within a company's constellation is different (usually assumed to be lower) than that of inter-constellation collision avoidance efficacy due to better coordination within an organization.

\section{Validation}
\label{sec:validation}
\subsection{Validation against the IADC study} 

The Inter-Agency Space Debris Coordination Committee (IADC) in 2009 had several space agencies use their MC tools to compare the performance for a strict future scenario \cite{iadccomparison}.  Several models contributed and were compared with each other: ASI: Space Debris Mitigation long-term analysis program (SDM), ESA: Debris Environment Long-Term Analysis Model (DELTA), ISRO: KS Canonical Propagation Model (KSCPROP), JAXA: LEO Debris Evolutionary Model (LEODEEM), NASA: LEO-to-GEO Environment Debris Model (LEGEND), UKSA: Debris Analysis and Monitoring Architecture for the Geosynchronous Environment (DAMAGE).  

A 2009 baseline environment for debris 10 cm and larger was provided by ESA.  The future space traffic model was based on a repetition of the historic 2001-2009 space traffic.  Each participating member used its own solar flux projection model. A catastrophic collision was defined as one characterized by an impacter's kinetic energy to the target mass ratio of 40 J/g or greater. A future post-mission disposal (PMD) compliance level of 90\% was assumed for both the spacecraft and launch vehicle stages. 
The initial simulation epoch was 2009 and was simulated for 200 years.  The initial object count was 17070 based on the MASTER database, which included payloads, derelict objects, rocket bodies, and debris.  Launches were repeated for launches between the years 2002 and 2008.  The payload lifetime was assumed to be 8 years and no collision maneuvers were assumed for any objects.  No explosions were assumed.  

With the identical input setup, the six models yielded similar qualitative results and confirmed the instability of the current LEO object population.  Despite an assumed global PMD level of 90\%, the six models resulted in a steady increase in the $>10$ cm population resulting in catastrophic collisions occurring every 5 to 10 years.  The majority of catastrophic collisions occurred near the 800 km and 1000 km altitudes due to high concentrations of space objects there.  The study concludes by noting that compliance with existing national and international space debris mitigation measures will not be sufficient to constrain the future population of LEO objects.  To stabilize the LEO environment, more aggressive measures, especially the removal of the more massive nonfunctional spacecraft and launch vehicle stages, should be considered and implemented in a cost-effective manner.

%%%%%%%%%%%%%%%%%

\begin{figure}[!ht]
    \centering    \includegraphics[width=0.5\textwidth]{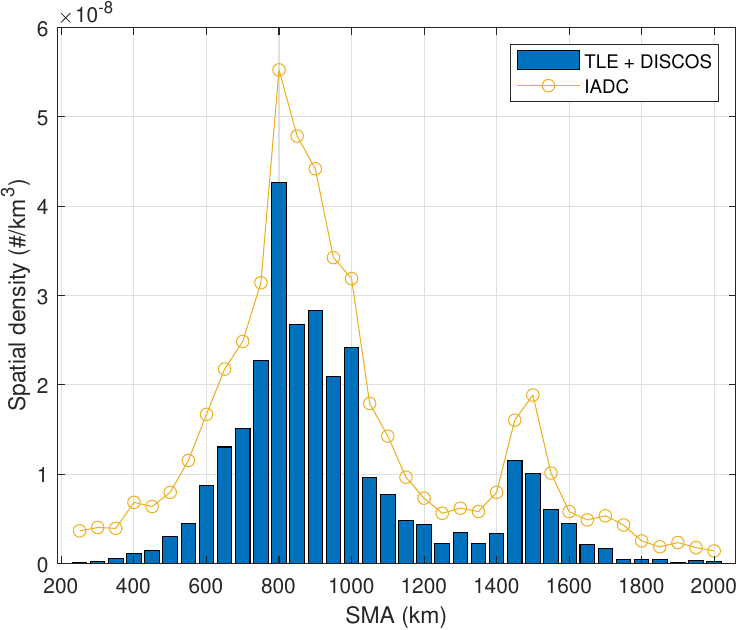}
    \caption{Spatial density of the initial population used in the IADC study vs TLE data for May 2009 epoch}
    \label{fig:iadcIPOP}
    % analysis_IADC2009.m
\end{figure}

MOCAT-MC was used to run the same scenario to validate its results against the six models used in the IADC study.  Although the specifics of the simulation data were not available, the best estimates were used to emulate the same scenario.  For example, the initial population of the IADC study starts with 17074 objects in LEO, though the TLEs available for the same epoch yielded around 9874 objects.  To start with the same number of objects with similar altitude density profile, a total of 7200 debris were added with the orbital and physical properties randomly selected from existing debris.  The altitude of the additional debris was selected to match the initial spatial density of the initial population provided in the study as seen in Fig. \ref{fig:iadcIPOP}.  Although the study was limited to trackable objects, the additional population was likely in the smaller objects compared to the tracked objects with orbits in the TLE catalog.  To account for this size discrepancy in debris sampling, the characteristic length of the expanded initial population of TLE/DISCOS was reduced by a factor of 1.5.  In addition, the PMD scheme assumed within MOCAT removes the object from the simulation when PMD is successful.

\begin{figure}[!ht]
    \centering
    \includegraphics[width=0.6\linewidth]{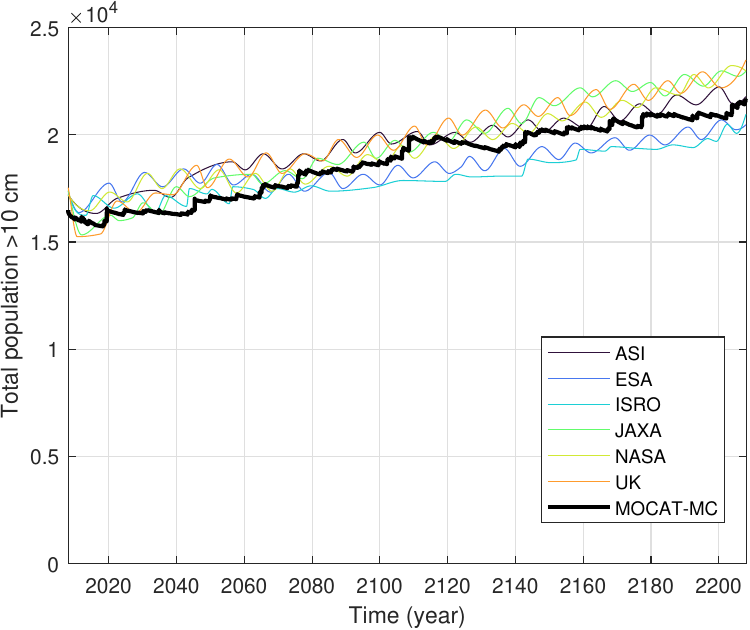}
    \caption{Comparison of total population between MOCAT-MC and IADC models}
    \label{fig:iadc-comparison}
    % analysis_0805.m
    % Validation_results/0805_historicNFL/iadc_plotDigitized.m
\end{figure}

\begin{figure}[!ht]
    \centering    \includegraphics[width=0.6\linewidth]{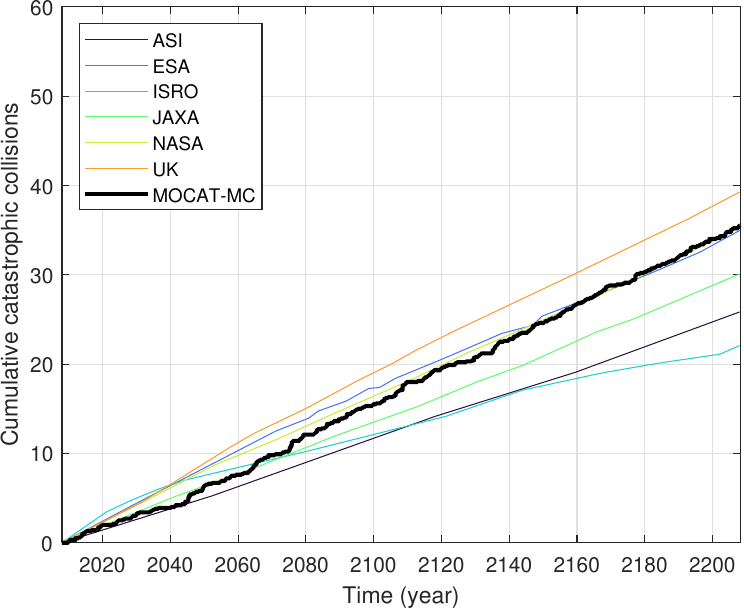}
    \caption{Comparison of cumulative catastrophic collisions between MOCAT-MC and IADC models}
    \label{fig:iadc-comparison2}
\end{figure}

\begin{figure}[!ht]
    \centering    \includegraphics[width=0.6\linewidth]{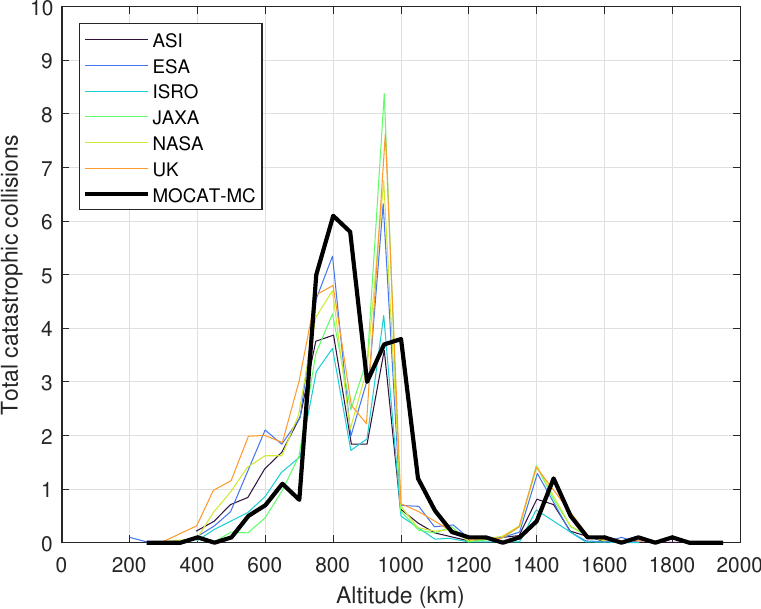}
    \caption{Altitude of catastrophic collisions between MOCAT-MC and IADC models between 2009-2209}
    \label{fig:iadc-comparison3}
\end{figure}

The summary of the MOCAT-MC validation exercise using the IADC study scenario is shown in Table \ref{tab:iadcVNVsummary}.  The average of 10 MC runs represent the MOCAT-MC results.  The comparison of the total population between MOCAT-MC and the models used in the IADC study is shown in Fig. \ref{fig:iadc-comparison}, and the cumulative catastrophic collisions and the altitude of catastrophic collisions are shown in Figs. \ref{fig:iadc-comparison2} and \ref{fig:iadc-comparison3}, respectively.  MOCAT-MC performed similarly to those of the IADC study for these key metrics.

\begin{table}[!ht]
\small
\ra{1.2}
\centering
\caption{Comparison between IADC study results and MOCAT-MC runs of the same NFL scenario}
\begin{adjustbox}{width=\textwidth}
\begin{tabular}{@{}lcccccccc@{}} %p{20mm}
\toprule
Agency                                                  & ASI   & ESA   & ISRO    & JAXA    & NASA   & UKSA   & All IADC & MIT  \\ %\midrule
Model                                                   & SDM   & DELTA & KSCPROP & LEODEEM & LEGEND & DAMAGE & - & MOCAT-MC    \\ \bottomrule
% \textbf{MC Runs}                                                 & 275   & 100   & 40      & 60      & 150    & 100    & 725   \\ \midrule
Runs with $N_{2209} > N_{2009}$ & 88\%  & 75\%  & 90\%    & 88\%    & 89\%   & 94\%   & 87\%  & 92\%\\ \midrule
Change in Population                    & +29\% & +22\% & +19\%   & +36\%   & +33\%  & +33\%  & +30\% & +35\% \\ \bottomrule
\end{tabular}
\label{tab:iadcVNVsummary}
\end{adjustbox}
\end{table}

\begin{figure}[!htb]
    \centering
    \includegraphics[width=0.6\linewidth]{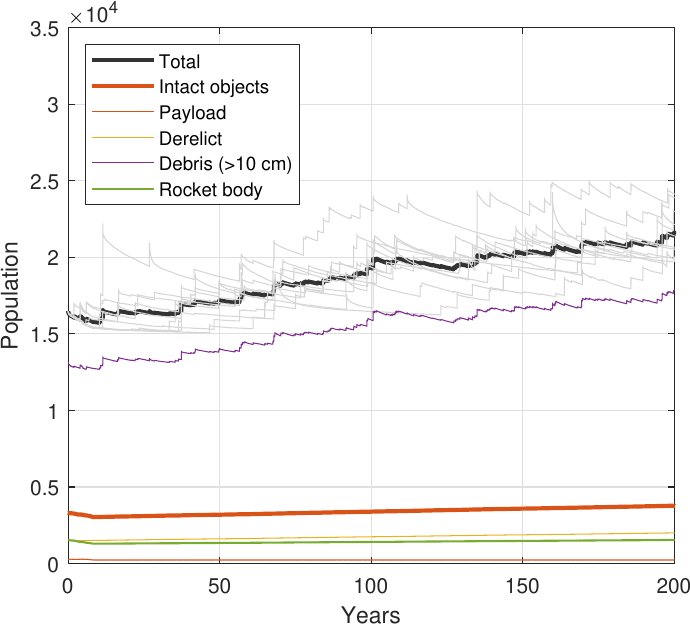}
    \caption{Population per object class for the IADC scenario from MOCAT-MC}
    \label{fig:iadc-mocat1}
    % thesisFigs/scripts/analyzeIADC.m
\end{figure}

\begin{figure}[!htb]
    \centering
    \includegraphics[width=0.6\linewidth]{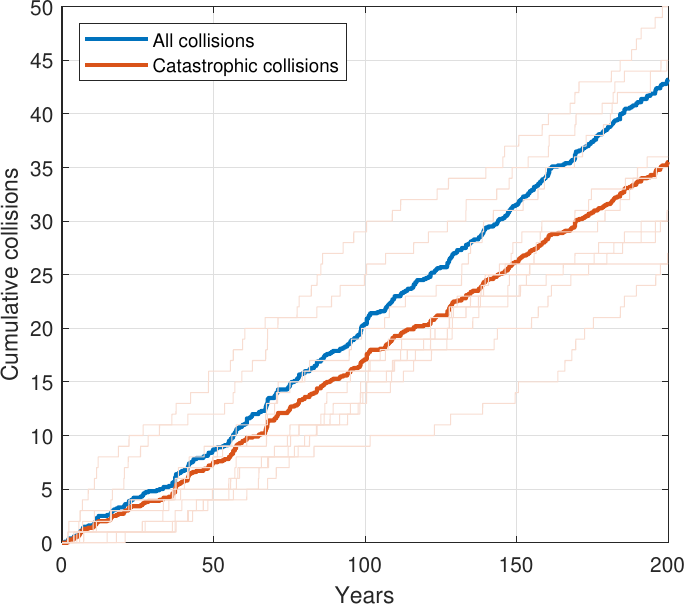}
    \caption{Cumulative collisions for the IADC scenario}
    \label{fig:iadc-mocat2}
    % thesisFigs/scripts/analyzeIADC.m
\end{figure}

The details of this MOCAT-MC validation scenario are shown in Fig. \ref{fig:iadc-mocat1}.  The models are run for objects that are $>10$ cm. The population per object class shows that the number of debris dominates the $>10$ cm population and contributes as the main source for the increase in the total object count.  The total population for some individual simulation is shown in gray for the population.  The intact objects (payload and rocket body) are relatively stable.  The collision statistics in Fig. \ref{fig:iadc-mocat2} shows the relative occurrence between catastrophic collisions and all collisions.  

The bottom portion shows the cumulative number of collisions, with the solid line denoting any collisions, whereas the dotted line shows only the catastrophic collisions.  In this scenario, most collisions are deemed catastrophic.

\section{Results and Discussion} \label{sec:results}
Many of the sub-functions were varied to understand the sensitivity of the simulation results on the various input parameters to the model.

\subsection{Benchmarking}

\begin{figure}[!ht]
    \centering	\includegraphics[width=0.5\linewidth]{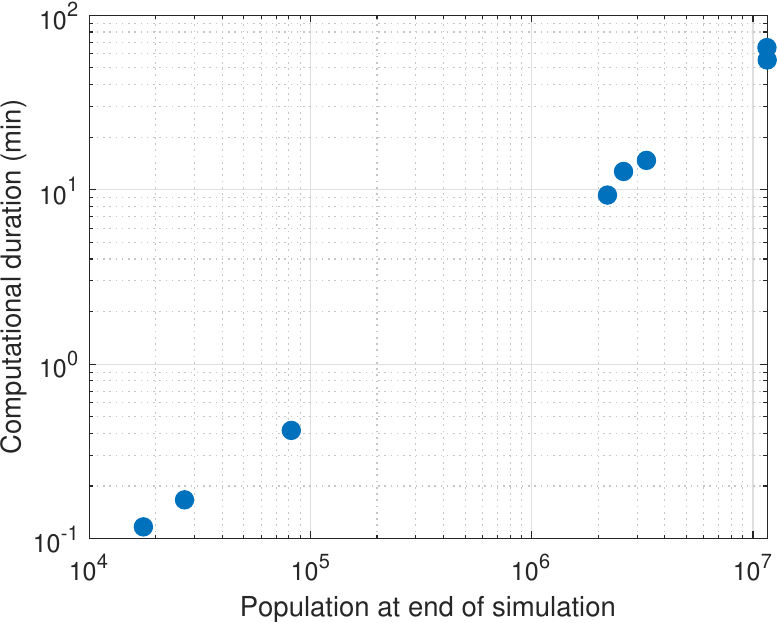}
	\caption{Run-time duration for MOCAT-MC}
	\label{fig:benchmark}
    % djangSBX/benchmarking_SSEMvsMC.m
\end{figure}

The computational resources required to run MOCAT-MC are shown in Fig. \ref{fig:benchmark} for a range of 100-year simulations on the SuperCloud system \cite{Reuther2018InteractiveAnalysis}, with Intel Xeon
Platinum 8260 processors.  Within MOCAT-MC, both the collision detection algorithm through Cube and the propagator complexity scale as $\mathcal{O}(n)$, which is shown in the computational duration.  Populations of 17000 to $10^7$ were run.  MOCAT-MC is a single-threaded simulator -- multiple threads and cores would allow for simultaneous runs of different MC simulations, but it will not inherently improve the run time for a particular MC simulation.  

\subsection{Extrapolated and No Future Launch cases with 2023 epoch}

Two scenarios are run with MOCAT-MC to compare the effect of future launches on the trackable orbital population in LEO with scenario epoch at Jan 1 2023:
\begin{itemize}
    \item Extrapolation of the recent launch traffic, explosion rates, and post-mission disposal rates
    \item No future launches scenario where no launches take place after 2022.
\end{itemize}

The TLE catalog is used as the orbital parameters of the trackable debris population with reference epoch Aug 1 2022, while the ESA DISCOS database was used for parameters such as the objects' size, mass, object type and launch date.  Only objects with perigee between 200 km and 2000 km were considered. The scenario parameters are described in Table \ref{tab:EXTvsNFLscenarios}.  The initial population profile is shown in the Appendix Table \ref{tab:TLEinitpop}. % and Fig. \ref{fig:TLEinitialSMA}.  

\begin{table}[ht!]
\centering
\caption{Scenario parameters}
% \begin{adjustbox}{width=\textwidth}
\begin{tabular}{@{}lccccccc@{}}
\toprule
Scenario & $T_{scenario}$ & $\Delta t$ & $a_p$  limit & Launch profile & $\mathbb{E}[P_{explosion}]$ & $P_{PMD}$ & PL lifetime \\ \midrule
NFL & 100 years & 5 days & $200 < a_p < 2000$ &  n/a  & n/a  & n/a & n/a \\ \midrule
Extrapolated & 100 years & 5 days & $200 < a_p < 2000$ & \makecell{repeat \\2017-2021} & RB: 2.3 / yr  & 0.4  & 8  years  \\ \bottomrule
\end{tabular}
% \end{adjustbox}
\label{tab:EXTvsNFLscenarios}
\end{table}

\begin{figure}[!htb]
    \centering
    \includegraphics[width=0.5\linewidth]{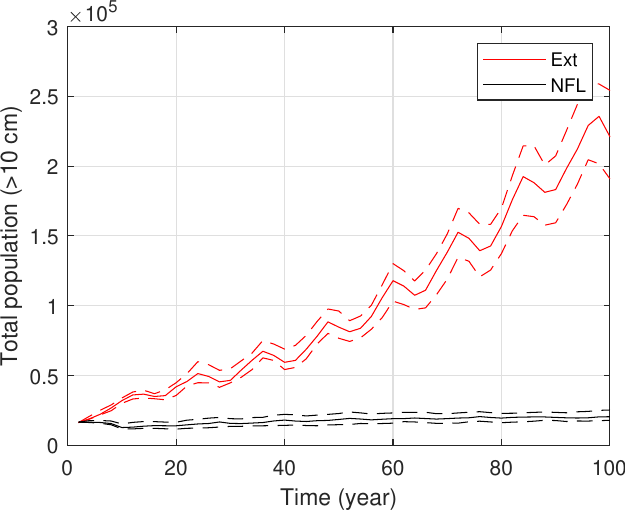}    
	% \subcaptionbox{Population}[.45\textwidth]{\includegraphics[width=0.9\linewidth]{Figures/ESA_ext_vs_nfl3.pdf}}
	% \subcaptionbox{Cumulative collisions}[.45\textwidth]{\includegraphics[width=0.9\linewidth]{Figures/ESA_ext_vs_nfl_collisions3.pdf}}  	\caption{Extrapolated and No Future Launch Scenarios for Epoch 2023}	
 \caption{Total population for the Extrapolated and No Future Launch scenarios with 2023 epoch}
 \label{fig:ESA_ExtNfl1}
    % \djangSBX\run_ESA_nfl_vs_ext_redux2024
    % \djangSBX\analyze_frag_out_dir.m
\end{figure}

\begin{figure}[!htb]
    \centering
    \includegraphics[width=0.5\linewidth]{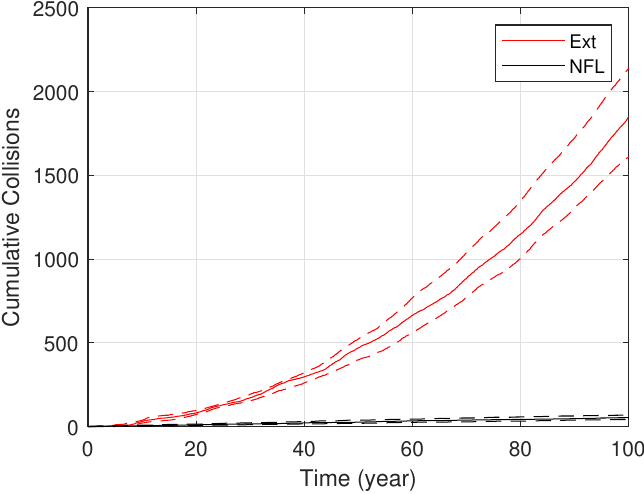}    
    \caption{Cumulative collisions for the Extrapolated and No Future Launch scenarios with 2023 epoch}
    \label{fig:ESA_ExtNfl2}
\end{figure}

All simulations were run for 100 years, and 100 Monte Carlo simulations were run on the MIT SuperCloud High Performance Computing cluster \cite{Reuther2018InteractiveAnalysis}.  
The explosion rate for rocket bodies and payload lifetimes is taken from the ESA Annual Report \cite{ESAspaceEnvReport2022}.  
There are 1594 objects from the TLE catalog that are not depicted in this initial population, as they have an SMA greater than $2000$ km, whose perigee is within the altitude limit.  Figure \ref{fig:ESA_ExtNfl1} shows the evolution of the total population of the simulation, and Fig. \ref{fig:ESA_ExtNfl2} shows the cumulative collisions over the 100-year simulations.  The dotted lines above and below denote the $3\sigma$ boundaries.  The growth of objects can be seen even for the NFL case.

It is difficult to compare the computational duration required by MC models, as they are not typically disclosed in the literature. 
Discussions with authors and researchers using these models reveal that MC simulations can take days to weeks \cite{LiouLEGENDlecture2012}.  However, a comparison of the maximum population modeled is disclosed in the published literature. Although not all published work aims to maximize the number of objects to model, some comparison and references are shown in Fig. \ref{fig:MCmaxpopLitreview}, which demonstrates the computational efficiency of MOCAT-MC. As shown in the earlier sections, simulations of more than millions of objects will be necessary to model the full proposed future megaconstellations.  Measuring the effect of lethal non-trackable objects will also require high population simulations. 

%% tikz plot comparing the literature:

\begin{figure}[!ht]
\centering
\begin{tikzpicture}
\begin{axis}[
    width = 12cm,
    height = 8cm,
    ytick={5,4,3,2,1},
    yticklabels={NASA LEGEND, UKSA DAMAGE, ESA DELTA, Others/Multiple, MOCAT-MC},
    xmode=log,
    log basis x={10},
    xlabel={Maximum Number of Objects in Simulation},
    ylabel={Model},
    xmin=5000, xmax=10000000,
    ymin=0.5, ymax=5.5,
    grid=both,
    minor tick num=1,
    grid style={line width=.1pt, draw=gray!10},
    major grid style={line width=.2pt,draw=gray!50},
    scatter/classes={%
        a={mark=square*, scale=1.5, fill=red,draw=black}, % Color for MOCAT
        b={mark=square*, scale=1.5, fill=blue,draw=black}, % Color for Others
        c={mark=square*, scale=1.5,fill=green,draw=black}, % Color for ESA
        d={mark=square*, scale=1.5,fill=orange,draw=black}, % Color for UKSA
        e={mark=square*, scale=1.5,fill=purple,draw=black} % Color for NASA 
    }
    ]
    \addplot[scatter,only marks,scatter src=explicit symbolic]table[meta=label] {
    x y label
    25000 1 a
    8500000 1 a
    9900000 1 a
    20000 2 b
    120000 2 b
    95000 2 b
    220000 2 b
    100000 3 c
    185000 3 c
    220000 3 c
    390000 3 c
    10000 4 d
    50000 4 d    
    72000 4 d
    49000 4 d
    25000 5 e
    20000 5 e
    55000 5 e
    75000 5 e
    };
    \node at (axis cs:25000,5) [anchor=south west] {\cite{LEGENDpmd2005}};
    \node at (axis cs:20000,5) [anchor=north west] {\cite{Liou2010ControllingRemoval}};
    \node at (axis cs:55000,5) [anchor=north west] {\cite{LEGENDpmd2013}};
    \node at (axis cs:75000,5) [anchor=south west] {\cite{Liou2011}};
    \node at (axis cs:10000,4) [anchor=north west] {\cite{LewisActiveDAMAGE}};
    \node at (axis cs:49000,4) [anchor=north west] {\cite{LewisUnderstandingDynamics}};    
    \node at (axis cs:72000,4) [anchor=north west] {\cite{Lewis2021DeepTime}};
    \node at (axis cs:49000,4) [anchor=south west] {\cite{White2014TheRemoval}};
    \node at (axis cs:100000,3) [anchor=north west] {\cite{Virgili2016}};
    \node at (axis cs:185000,3) [anchor=north west] {\cite{DELTA2}};
    \node at (axis cs:220000,3) [anchor=south west] {\cite{ESAspaceEnvReport2023}};
    \node at (axis cs:390000,3) [anchor=north west] {\cite{2023Letizia}};
    \node at (axis cs:220000,2) [anchor=north west] {\cite{LewisSensitivity}};
    \node at (axis cs:20000,2) [anchor=north west] {\cite{iadccomparison}};
    \node at (axis cs:120000,2) [anchor=south west] {\cite{iadc2016riskToSpaceSustain}};
    \node at (axis cs:95000,2) [anchor=north west] {\cite{Giudici2024}};
    \node at (axis cs:25000,1) [anchor=north west] {\cite{MOCAT-AAS-2022}};
    \node at (axis cs:8500000,1) [anchor=north east] {\cite{MITRI}};
    \node at (axis cs:9900000,1) [anchor=south east] {\cite{MOCAT-bin-AMOS2022}};
\end{axis}
\end{tikzpicture}
\caption{Comparison of population modeled in existing MC simulation tools}
\label{fig:MCmaxpopLitreview}
\end{figure}

\subsection{No Future Launch Cases from the Past}

The No Future Launch case shown in the Validation section shows growth in the number of objects and in the number of collisions despite no new launches occurring.  The literature has shown that MC methods point to the fact that the LEO environment was in an era of unabated growth for decades prior \cite{ESAspaceEnvReport2022, LEGEND2004, DELTA2}. 
Using the same methodology as in the previous section with the combined dataset between the TLE catalog and the DISCOS database, the future LEO environment is simulated at different epochs in the past.  The NFL case is run for every year starting in the year 2000 with the cataloged objects at those epochs and removing all future launches to see the growth or decay in the number of objects. 

January 1 of each scenario epoch year is used for the initial population of objects for the LEO objects, which inherently limited the initial population to the tracked objects. All payloads were assumed to have a mission lifetime of 8 years, after which they undergo PMD with 90\% success rate.  Collision avoidance failure rate of $\alpha = 0.01$ and cube resolution of 10 km.  No new objects were introduced into the environment other than through collision dynamics, and no explosions were modeled.  The simulations were run for 100 years.  

\begin{figure}[!hbt]
    \centering
    \includegraphics[width=0.6\textwidth]{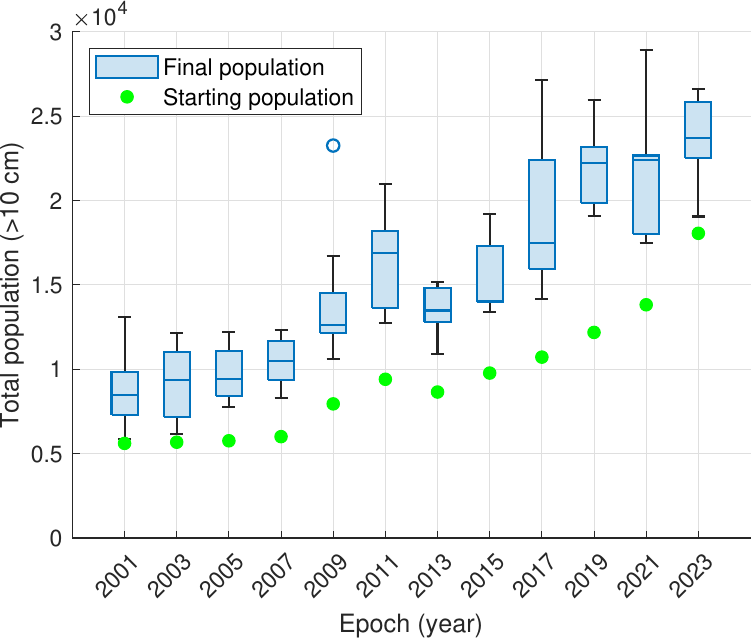}
    \caption{Object count for NFL cases starting at different starting epochs simluated for 100 years}
    \label{fig:historicalNFL}
    % Validation_results\0805_historicNFL\analysis_0805.m
\end{figure}

\begin{figure}[!hbt]
    \centering
    \includegraphics[width=0.6\textwidth]{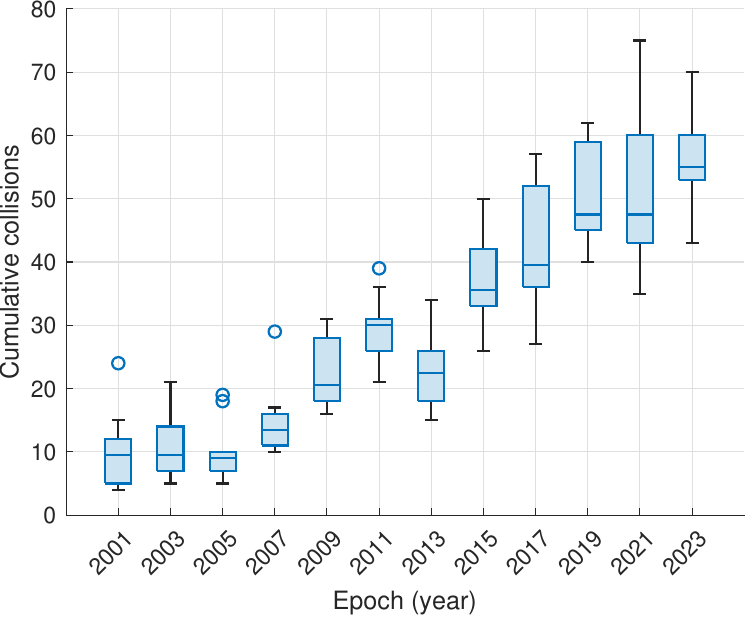}
    \caption{Total cumulative collisions for NFL cases starting at different starting epochs simluated for 100 years}
    \label{fig:historicalNFLcollisions}
\end{figure}

The results are summarized in Figs. \labelcref{fig:historicalNFL,fig:historicalNFLcollisions}. All of the simulations show a substantial growth from the initial population, which indicates that even without future launches, the unabated growth of objects is seen at least as far back as the year 2000 when there were 5000 tracked objects in orbit.

The results show that the growth in the number of objects is consistently present even with epoch in the year 2000, showing the importance of debris mitigation with any population.  Of note are the more recent years, where the total count of trackable objects remain roughly steady compared to previous years despite the higher initial population count; in these scenarios, the Starlink satellites have begun to populate, taking up a significant portion of the initial population for those epochs.  These active satellites will have been removed entirely from the environment within 8 years of the epoch, other than the few that remain as derelict satellites after failed PMDs.  Compared to other satellites, the average lifetime for these active payloads at the simulation epoch are much shorter; therefore, the effect on the space environment is much less.  For two scenarios with the same number of population, the one with a greater proportion of active satellites that can CAM and PMD will yield a safer and sustainable long-term LEO environment.

\begin{figure}[ht!]
    \centering
    \includegraphics[width=0.5\textwidth]{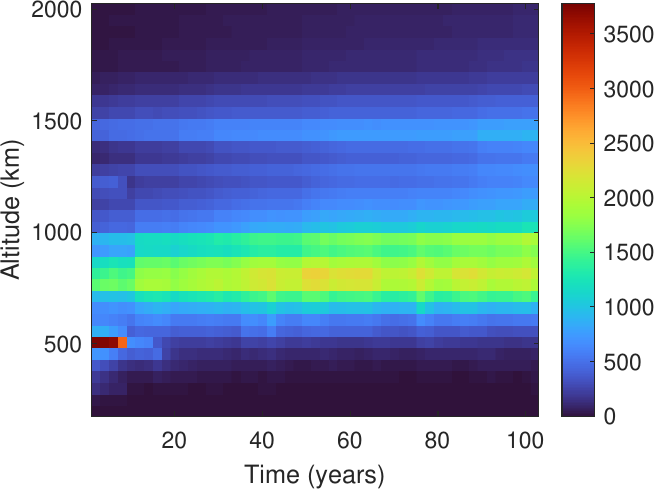}
    \caption{Total population (>10 cm) per altitude for the NFL case with 2023 epoch}    \label{fig:nfl2023}
    % analysis_0805.m
\end{figure}

\begin{figure}[!htb]
    \centering
    \includegraphics[width=0.5\linewidth]{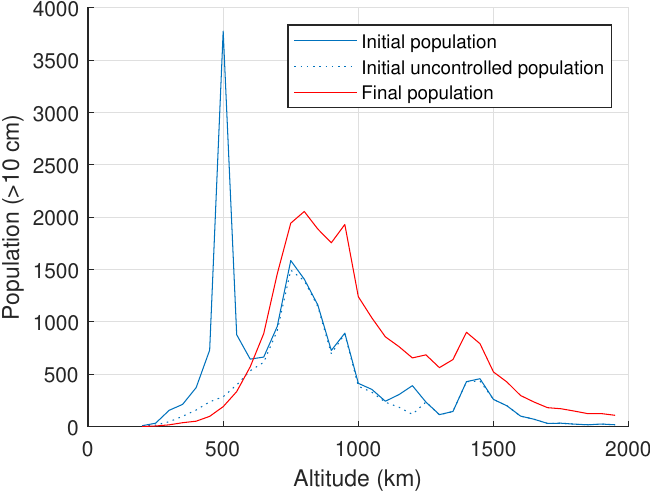}
    \caption{Initial and final population for NFL case with 2023 epoch}
	\label{fig:nfl2023colAlt1}
    % analysis_0805.m
\end{figure}

\begin{figure}[!htb]
    \centering
    \includegraphics[width=0.5\linewidth]{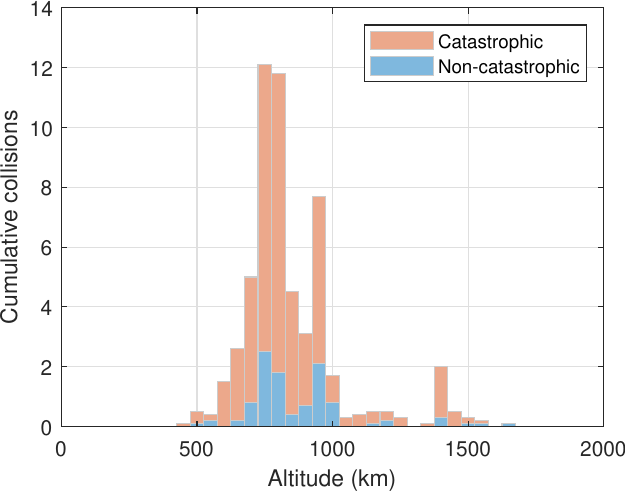}
    \caption{Cumulative collision per altitude for NFL case with 2023 epoch}
	\label{fig:nfl2023colAlt2}
    % analysis_0805.m
\end{figure}

The population divided into altitude is shown for the 2023 NFL case in Fig. \ref{fig:nfl2023}.  In this figure, the altitude is binned at 50 km and the time is binned yearly. The dense region in the first few years around 550 km denotes the numerous Starlink satellites which is mostly removed after its mission lifetime through  successful PMD.  This is also seen in the payload population in the Appendix. To a lesser degree, the OneWeb constellation can be seen around 1200 km band, which also goes through a similar drop in population after its mission lifetime.  % Fig. \ref{fig:appendix-historicalNFL-S}.  
The number of objects above the 700 km range continues to grow, though most objects below deorbit relatively quickly.  The growth in the 1000 km and 1400 km altitude region is seen, where many derelicts and debris exist and are relatively unaffected by atmospheric drag.  The altitude regions with higher density start to merge as collisions increase the number of objects and add objects to a range of altitudes around the collision.  The difference between the initial and final population is more pronounced in the comparison shown in Fig. \ref{fig:nfl2023colAlt1}.  The cumulative collision occurrence is shown in Fig. \ref{fig:nfl2023colAlt2}.  In these figures, the altitude is binned at 50 km.

\subsection{Future Traffic with Megaconstellations} \label{sec:Megaconstellations}
Megaconstellations are large networks of interconnected satellites that are designed to provide various services, including global broadband internet coverage, Earth observation, and communication capabilities. Two prominent examples of proposed megaconstellations are SpaceX's Starlink and OneWeb.  The race to deploy megaconstellations is driven by the potential benefits of widespread internet access, improved communication infrastructure, and enhanced Earth observation capabilities.

It is impossible to forecast the exact state of LEO traffic and launches in the upcoming decades.  The best estimate of megaconstellation data can be compiled from various sources, particularly governmental and regulatory bodies such as the FCC and ITU as well as press releases.  As of writing, there are more than 50 megaconstellations -- defined as constellations comprising more than 1000 satellites -- that have been credibly proposed \cite{mcdowell2023jonathan,henry2019spacex,diaz2023data}.  Of these, only a few have operational satellites at the time of writing.  As these are all commercial ventures, the technology roadmap, market conditions, and consumer demand make the forecasting of successful launch and operation of megaconstellations difficult. 

Some examples of megaconstellations are described below. These examples highlight the growing interest in megaconstellation projects across various countries and industries, notably in the internet connectivity and remote sensing applications.  Even in the past few years, many prominent megaconstellation projects that have filed with the ITU and/or FCC have been canceled or merged with other efforts.  Future launches and traffic due to megaconstellation should be taken as estimates at best; however, this list gives a general overview of launches to expect.  

\textbf{SpaceX \textit{Starlink}} \footnote{\url{https://www.starlink.com/}}
SpaceX initiated the Starlink project in 2015 with the goal of creating a satellite network capable of delivering high-speed, low-latency internet access worldwide. The primary motivation behind Starlink was to bridge the digital divide and provide reliable internet connectivity to underserved regions. The constellation would consist of thousands of small, low-Earth orbit (LEO) satellites that create a mesh network.
SpaceX began launching Starlink satellites in batches starting in May 2019.  As of September 2021, SpaceX had already deployed thousands of satellites, and beta testing of the Starlink service had begun in select regions.

\textbf{Eutelsat \textit{OneWeb}}  \footnote{\url{http://oneweb.net/}}
OneWeb is another megaconstellation project that aims to provide global broadband internet coverage. Founded in 2012, OneWeb intended to build a network of LEO satellites that could deliver internet services to remote and underserved areas. The project was supported by several notable investors, including SoftBank, Qualcomm, and the Government of the United Kingdom.  OneWeb faced financial challenges and had to file for bankruptcy in March 2020. However, the company was subsequently acquired by a consortium consisting of the British government and the Indian company Bharti Global. This acquisition provided the necessary funding to continue the project. OneWeb resumed satellite launches in December 2020 and has since progressed with its deployment plans.

\textbf{Amazon \textit{Project Kuiper}} \footnote{\url{https://www.aboutamazon.com/projectkuiper}}
Project Kuiper is Amazon's venture to create a megaconstellation of satellites to provide global broadband internet coverage. Announced in 2019, the project aims to deploy a network of more than 3200 LEO satellites. Like other megaconstellations, Project Kuiper's objective is to deliver affordable, low-latency internet services.

\textbf{Telesat \textit{Lightspeed}}  \footnote{\url{https://www.telesat.com/leo-satellites/}}
Telesat, a Canadian satellite operator, is working on its own megaconstellation known as Telesat Lightspeed. The project aims to provide broadband connectivity worldwide using a network of approximately 300 LEO satellites. Telesat's focus is on serving both residential and commercial markets, offering high-speed internet access, enterprise connectivity, and government services.

\textbf{E-space \textit{Semaphore-C}}  \footnote{\url{https://www.e-space.com/}}
Greg Wyler's company E-space made headlines when it registered 327000 satellites using Rwanda as the registration authority through the ITU in 2021.  In June 2023, the company filed another constellation \textit{Semaphore-C}, which is a constellation of 116540 satellites orbiting between 414 and 600 km altitude, registered in France.  Due to the recency of this addition, E-Space's constellation will not be part of this analysis.  Note that numerous other studies and the literature have ignored this constellation due to the perceived lack of credibility that the constellation will launch in its entirety.  The proposed constellation of more than 400,000 satellites would dwarf the total number of other proposed megaconstellations.  This highlights the academic and industry sentiment that regulatory filings are necessary but not sufficient to be used as credible sources for future LEO traffic.  

\begin{figure}[!ht]
    \centering    \includegraphics[width = .5\textwidth]{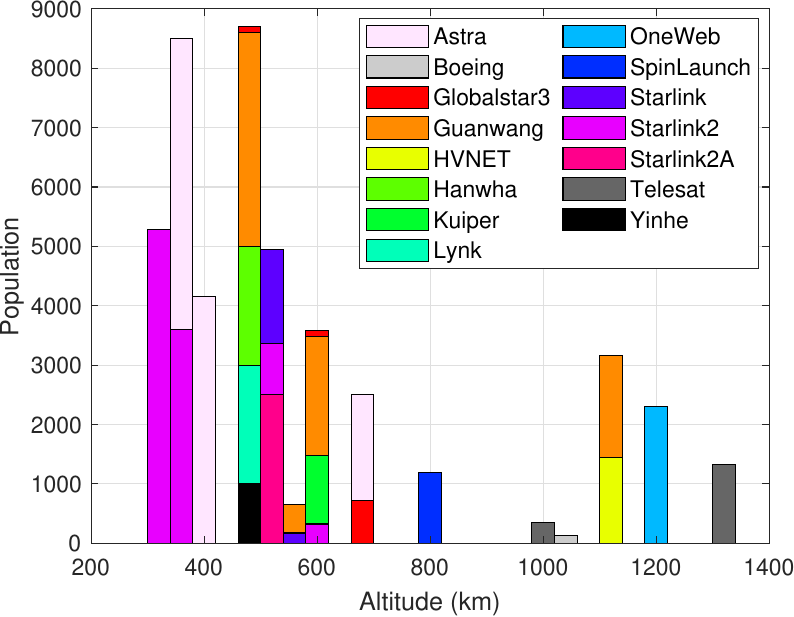}
    \caption{Modeled total operational population per megaconstellation}
    \label{fig:megaLaunch}
    %  Validation_results\0730_megasweep\analysis0730.m
\end{figure}

The future launch model used in this scenario is listed in the Appendix Table \ref{tab:megalaunch}, and totals more than 82000 satellites in operation just from the megaconstellations alone.  The missing mass and radius represented with a `-' used the Starlink satellite as the surrogate (260 kg and 2 m radius), while the start and finish years were set to be the latest dates for the other constellations with some data (2035 to 2055 for Guanwang).  Figure \ref{fig:megaLaunch} visualizes the launches.  Note that the replenishment launches are not shown in this figure.  An average of $n/yr_\text{missionlife}$ satellites per year will be required to launch in order to maintain the current number of satellites $n$. 

Three subsets of these megaconstellation launch scenarios are also chosen to be simulated. 
\textit{Case 1}: all of the filed megaconstellations as shown in Fig. \ref{fig:megaLaunch}, which totals 84139 operational megaconstellation satellites. 
\textit{Case 2}: just the constellations filed by Starlink, Kuiper, and OneWeb, all of which have some constellation presence as of Jan 2024.  This totals 44716 operational megaconstellation satellites.
\textit{Case 3}: which comprises of just Starlink v1, v1.5, Kuiper and OneWeb.  Compared to Case 2, this case removes the largest megaconstellation proposed considered, which is Starlink v2.  This totals 22228 megaconstellation satellites.

\begin{figure}[!ht]
    \centering
    \includegraphics[width=0.6\textwidth]{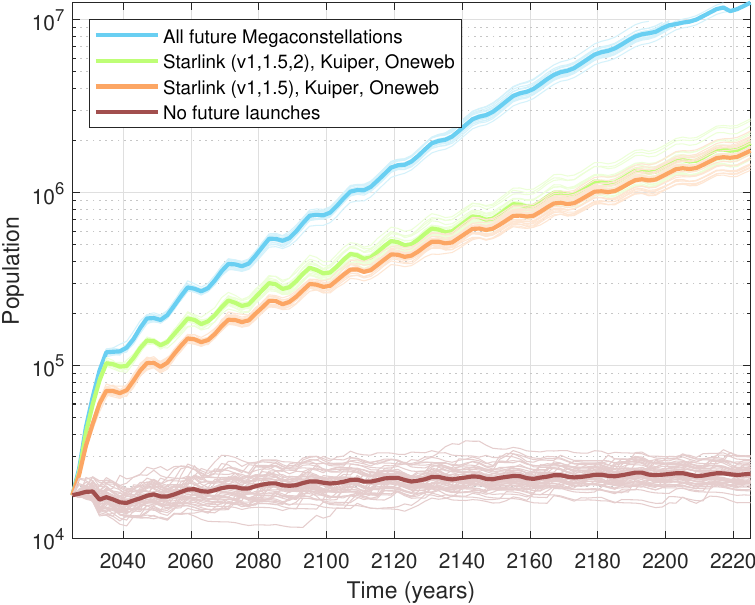}
    \caption{Total population (>10 cm) in LEO with future megaconstellation launches}
    \label{fig:megaprop}
    % Validation_results\0730_megasweep\analysis.m
    % megapropAMOS*.fig
\end{figure}

\begin{figure}[!h] % \label{fig:megaprop_S} \label{fig:megaprop_D}
    \centering
	\subcaptionbox{Payload population}[.4\textwidth]{\includegraphics[width=0.9\linewidth]{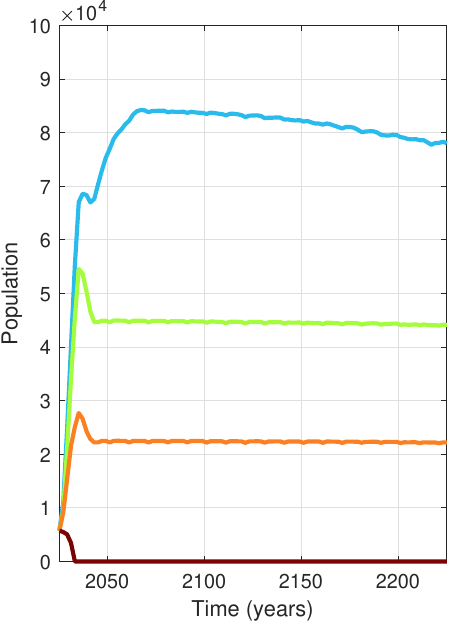}}	
	\subcaptionbox{Derelict population}[.4\textwidth]{\includegraphics[width=0.9\linewidth]{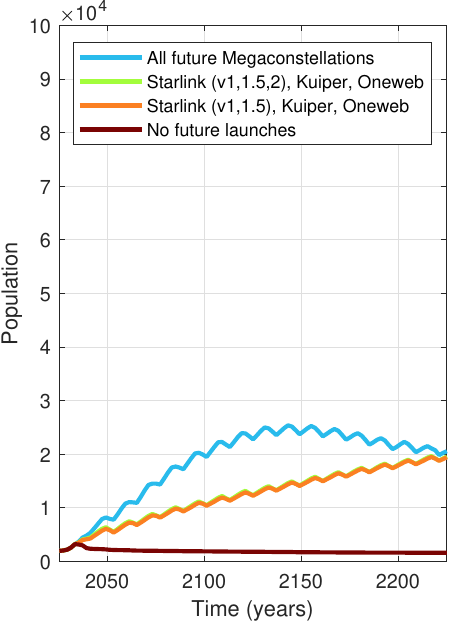}}
    \caption{Payload and derelict population with various future megaconstellation launch models}
    \label{fig:megaprop_S-D}
    % Validation_results\0730_megasweep\analysis0730.m
    % megapropAMOS*.fig
\end{figure}

\begin{figure}[!h]
    \centering
    \includegraphics[width=0.6\linewidth]{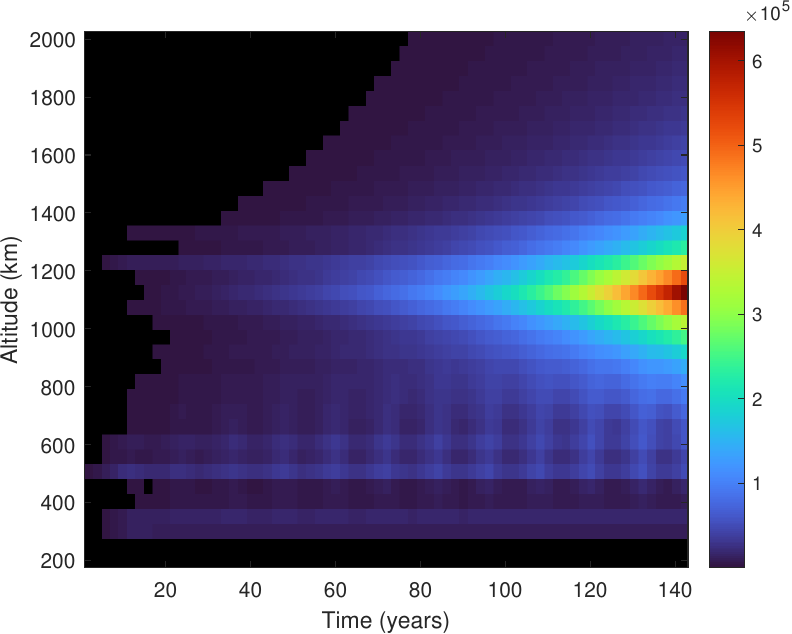}
    \caption{Total population (>10 cm) per altitude for the all future megaconstellations case}
    \label{fig:mega1}
\end{figure}

The simulation results for the three cases along with the case of no future launches are shown in Fig. \ref{fig:megaprop}.  The growth in the number of objects is clearly seen in all cases.  Notable are the S (payload) and D (derelict) plots in Fig. \ref{fig:megaprop_S-D} – the number of objects increases with increasing launch scenarios, but at the highest number of launches, the number of payloads starts to decrease, despite the 1\% probability of failure to avoid a collision.  That effect is more pronounced in the derelict class, where such avoidance is not possible, and the number of derelict objects starts to dwindle after 100 years.  The initial decrease in the payload population around year 2030 is due to the number of existing objects in the initial population that are deorbiting after the assumed 8-year mission lifetime.  

In these figures, it is also seen that the number of objects is likely to grow even without new launches.  This is in line with the findings from the literature and highlights the urgent need to limit the creation of derelict objects through higher PMD rates and effective collision avoidance maneuvers to limit fragmentation events.  The temporal evolution of the population per altitude is shown in Fig. \ref{fig:mega1}.  Altitude is binned at 50 km and time is binned yearly.  The result clearly shows the large amount of accumulation above 1000 km region.  

\begin{figure}[!h]
    \centering
    \includegraphics[width=0.55\linewidth]{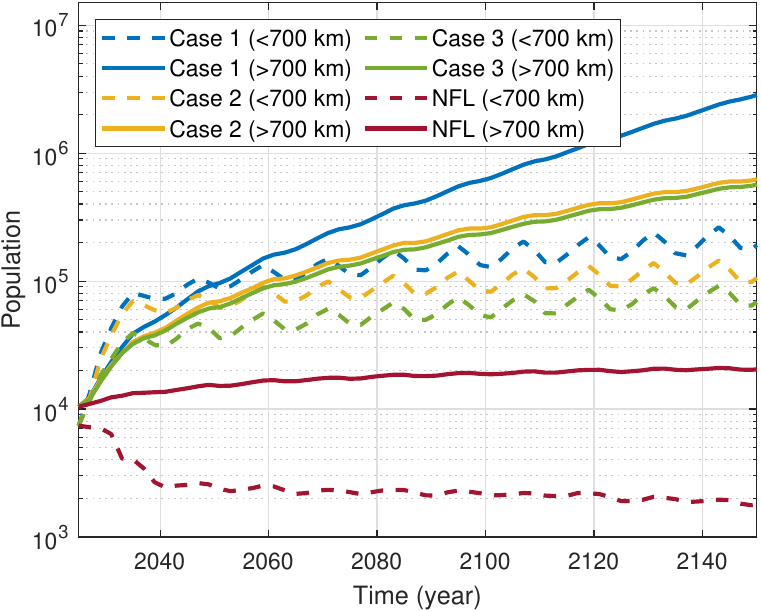}
    \caption{Total population (>10 cm) below and above 700 km altitude for the four future launch cases}
    \label{fig:mega700}
\end{figure}

The population difference between objects above and below 700 km shows the accumulation rate between the higher and lower altitudes.  Figure \ref{fig:mega700} shows the comparison for each of the four cases, with the dashed line denoting the population below 700 km and the solid lines denoting the population above 700 km.  Despite much higher launches occurring below 700 km, the population below 700 km remains largely steady and low, whereas the higher altitudes grow continuously.  This is pronounced even for the \textit{No Future Launch} case.  The lower altitude also exhibits undulation of the population as a result of the thermospheric expansion and contraction that follow the solar cycle.  The comparison of launching into only lower-altitude shells is explored in the next section.

\subsection{Megaconstellation launches with limited altitudes}
To see the effect of megaconstellation launches at higher altitudes, a subset of the total megaconstellation case was launched and analyzed.  The altitude limits of $<600$ km (total 59336 operational), $<700$ km (65408), $<900$ km (66598) were chosen.  The launch altitudes of the megaconstellations can be found in Fig. \ref{fig:megaLaunch}.  The same parameters as in the previous section were used, except for the launched subset.

\begin{figure}[!ht]
    \centering
	\subcaptionbox{Total population (>10 cm)}[.45\textwidth]{\includegraphics[width=0.9\linewidth]{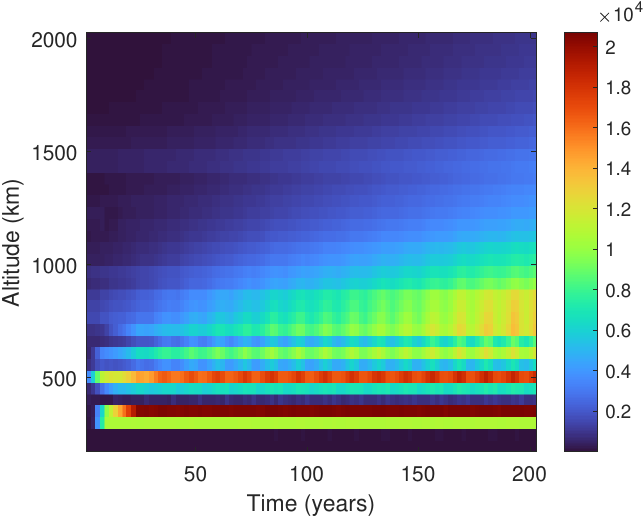}}
	\subcaptionbox{Collisions}[.45\textwidth]{\includegraphics[width=0.9\linewidth]{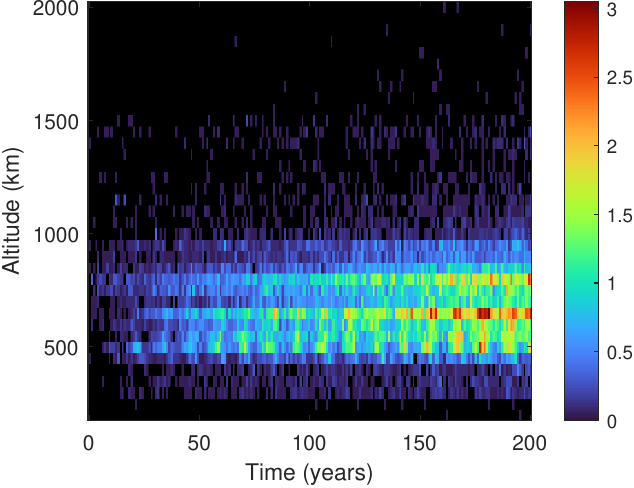}}  
	\caption{Megaconstellation launches limited to $<700$ km}
	\label{fig:megaAltLim700}
    % 0730_megasweep/analysis.m
\end{figure}

The collision and population statistics from the $<700$ km constellation case is shown in Fig. \ref{fig:megaAltLim700}, while the other cases are shown in the Appendix.   The comparison between these altitude-limited constellations is shown in the Appendix. %Fig. \ref{fig:megaAltLim600vs900}.  
It is clearly seen that the growth of the debris population and the collision rates are dominated by higher orbits.  Despite the much fewer objects launched $>700$ km, the total collision number across the 200-year simulation is more than halved.  
Note that for this analysis, the same PMD efficacy rate is used for all constellations. That is, the frequency of successful removal from the environment after the mission life of the active payload is the same for all payloads of the same type regardless of the altitude of the constellation.  The more numerous population for the $>700$ km launch scenario is due to the derelict satellites in higher orbits that remain in the environment for much longer due to lower atmospheric drag.  

\begin{figure}[!ht]
    \centering
	\subcaptionbox{Total population (>10 cm)}[.45\textwidth]{\includegraphics[width=0.9\linewidth]{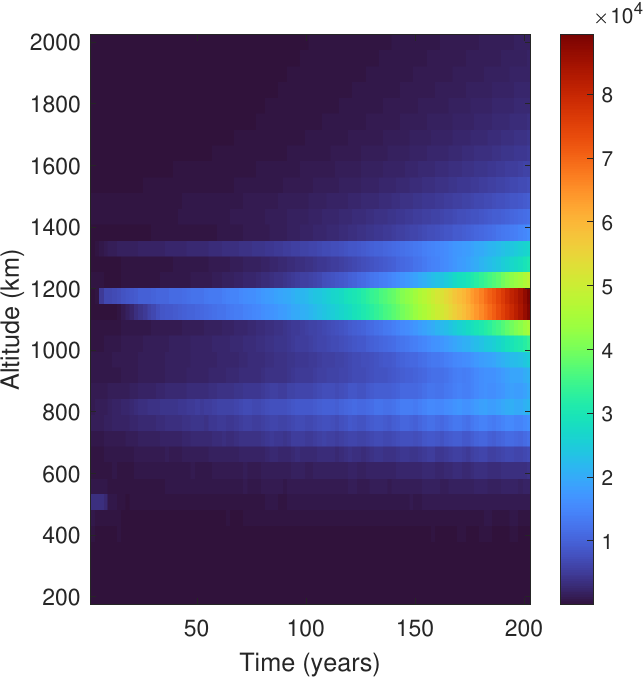}}
	\subcaptionbox{Comparison between $<700$ km and $>700$ km launch cases after 200 years of propagation}[.37\textwidth]{\includegraphics[width=0.95\linewidth]{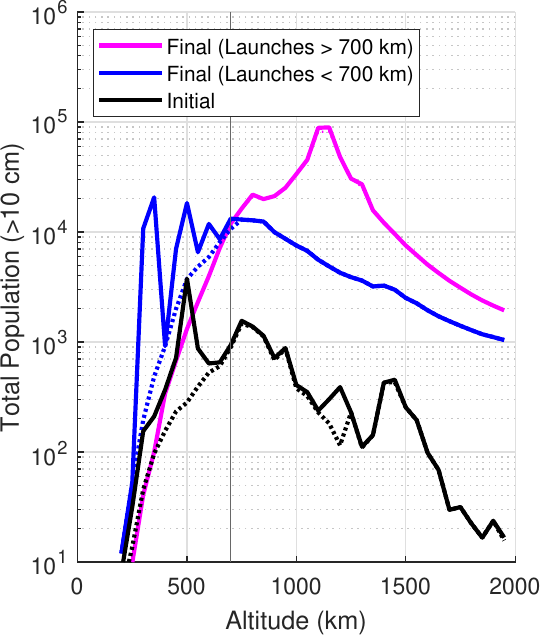}}  
	\caption{Megaconstellation launches limited to $>700$ km}
	\label{fig:megaAltLimAbove700}
    % 0730_megasweep/...
\end{figure}

For comparison, Fig. \ref{fig:megaAltLimAbove700} shows the population when only the $>700$ km constellations are launched. The altitude is binned at 50 km.  Although the operational satellites for the $>700$ km megaconstellations are about 1/4 of the population of the $<700$ km case, the population growth is much more pronounced. The reduced drag effect is clearly seen, with the debris population from the 800 km region persisting throughout the simulation duration.  In Fig. \ref{fig:megaAltLimAbove700}(b), the dotted lines denote the uncontrolled objects, including derelict and debris objects.  The solid line is the total number of objects.  The difference between the solid and the dotted line is the controlled payload population, where Starlink's contribution to the current population (Initial population) is clear around 500 km.  This comparison shows the relative difference between the debris population and the payload population.  The low-altitude constellations are able to have a much higher payload-to-debris ratio, enjoying a lower debris environment due to the atmosphere while also lowering the collision avoidance operational burden.  Higher orbits, despite having the same PMD and $\alpha$ collision avoidance efficacy, exist with a much higher debris-to-payload ratio as the debris accumulates.  This shows that debris mitigation efforts through higher PMD and effective collision avoidance will be crucial to maintaining a viable orbit regime for higher altitudes.

\section{Conclusions} \label{sec:conclusion}
As access to space is becoming easier and with the accelerated pace of launches into LEO, it is imperative to be able to model the future space environment and understand the various inputs that can shape the future.  This paper describes a novel open-source Monte Carlo-based method to simulate the evolution of the LEO population called MOCAT-MC.  The tool efficiently models the evolution of the orbital population characterized by dynamics such as launches, collisions, explosions, deorbit methods, and more.  This Monte Carlo tool is flexible in its modeling fidelity with several options for the propagator, initial simulation population, and launch profiles.  A sampling-based collision model is used and its sensitivity to input parameters is explored.  Statistical convergence is tested, and the output result has been validated against six other models in the literature with which MOCAT-MC shows good agreement. Additionally, MOCAT-MC is able to simulate up to 20 million objects over a period of 200 years, improving on the previous state-of-the-art, which was 400K objects over the same period. Historical look at the no-future launch cases shows that the LEO population without any new launches since 2000 may have continued to grow and has only accelerated with the increased population since then.  
A future launch case involving all megaconstellations filed with the ITU and FCC is explored, where the number of objects $>10$ cm grows to tens of millions of objects.  These results underscore the importance of a multitude of debris mitigation strategies that include international policies and technological solutions.  
The simulation results show some important conclusions: the higher--altitude accumulation of orbital debris is much faster than that of the lower altitudes and warrants careful planning.  In addition, the accumulation of debris in the higher orbits is not only affected by payload launches into the lower altitudes; collisions in any orbital regime will deposit debris into any other orbital altitudes due to the $\Delta V$ imparted during the fragmentation event.  Despite the much fewer megaconstellations planned at the higher altitudes, their failure in post-mission disposal or collision avoidance maneuvers will result in a large effect on orbital debris accumulation.  

% \clearpage

\section{Appendix}

\subsection{NASA SBM implementation}
\begin{figure}[!hb]
    \centering
    \includegraphics[width=0.7 \columnwidth]{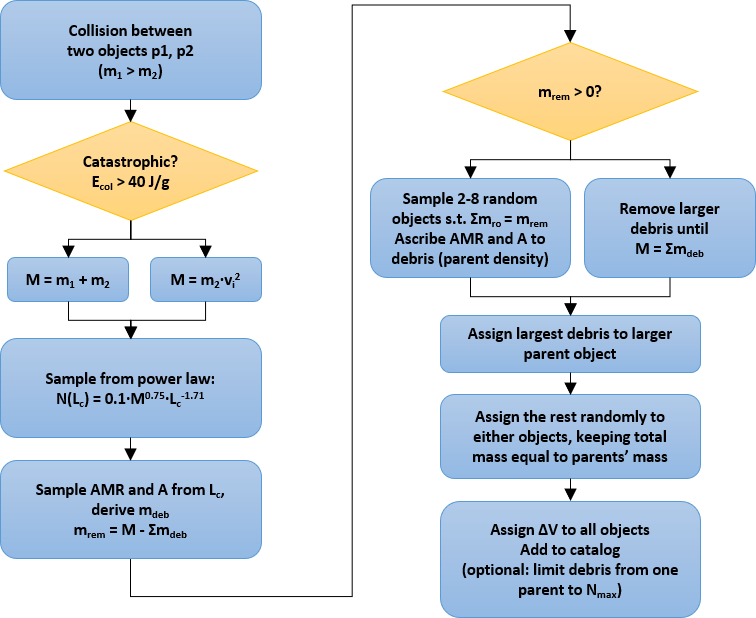}
    \caption{NASA Standard Breakup Model as implemented in MOCAT-MC}
    \label{fig:sbm_flow}
\end{figure}

The implementation of the NASA SBM in MOCAT-MC is shown in Fig. \ref{fig:sbm_flow}. 

\subsection{Variability of B* in TLEs}
\begin{figure}[!htb]
    \centering
    \includegraphics[width=0.5\textwidth]{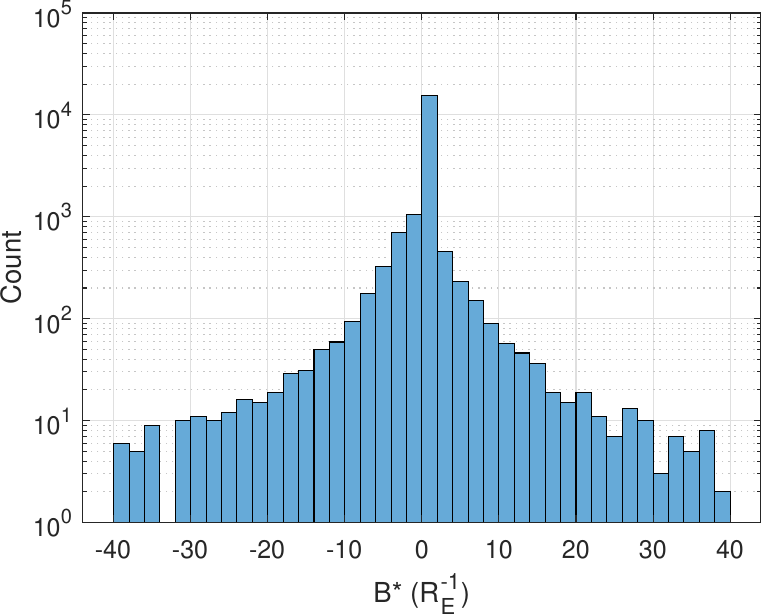}
    \caption{Distribution $B^*$ values from the TLE catalog (Jan 2023)}
    \label{fig:BstarDistribution2}
    % bstar_CD_comparison.m
\end{figure}

A snapshot of the $B^*$ values of the TLEs from January 2023 is shown in Fig. \ref{fig:BstarDistribution2}. The $B^*$ value is a fitted parameter, hence the non-physical negative values for a large number of objects.

\subsection{No Future Launch scenarios}

The data used for the \textit{No Future Launches} and \textit{Extrapolated} scenarios with January 2023 epoch are described here.  At this epoch, a large portion of the active payloads are from Starlink in the 500 km altitude bin.  Table \ref{tab:TLEinitpop} shows the breakdown of each object as described by DISCOS, where MRO represents Mission Related Objects, FD represents Fragmentation Debris, and D represents Debris as defined by ESA \cite{ESAspaceEnvReport2022}.  

\begin{figure}[!ht] 
    \centering
	\subcaptionbox{Payload}[.3\textwidth]{\includegraphics[width=0.95\linewidth]{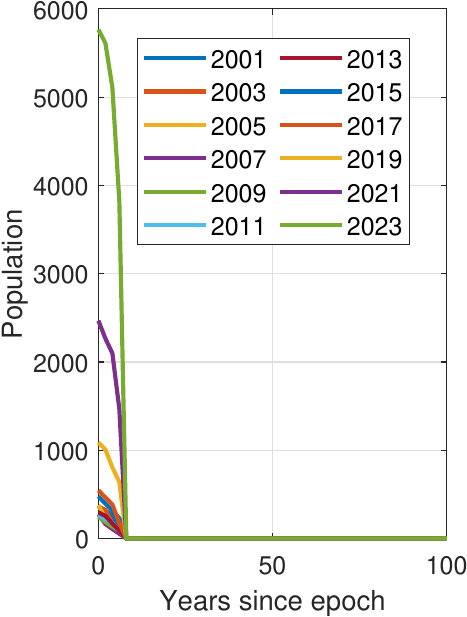}}	
	\subcaptionbox{Derelict}[.3\textwidth]{\includegraphics[width=0.95\linewidth]{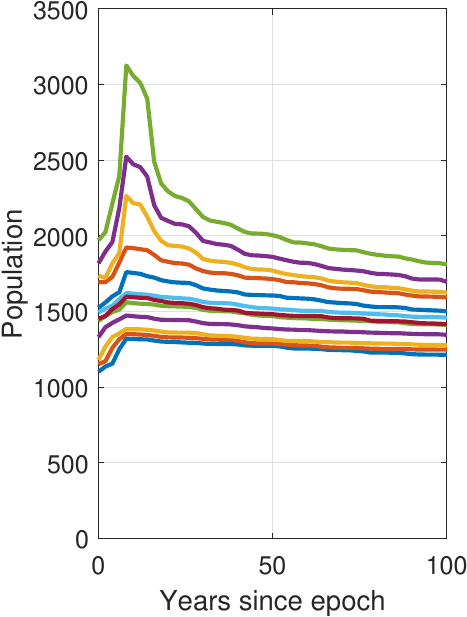}}		  
    \\ 
	\subcaptionbox{Debris (>10 cm)}[.3\textwidth]{\includegraphics[width=0.95\linewidth]{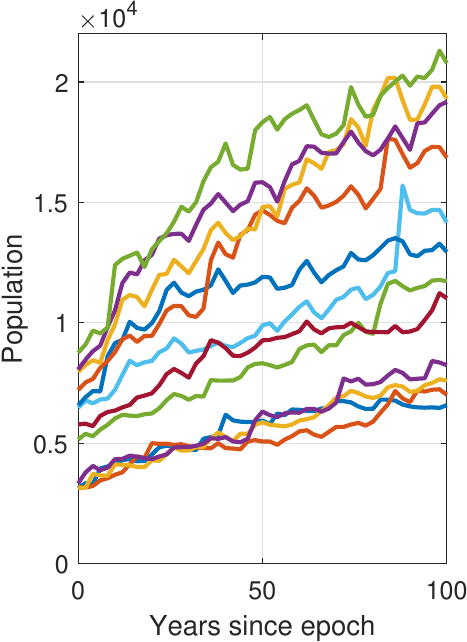}}	
	\subcaptionbox{Rocket body}[.3\textwidth]{\includegraphics[width=0.95\linewidth]{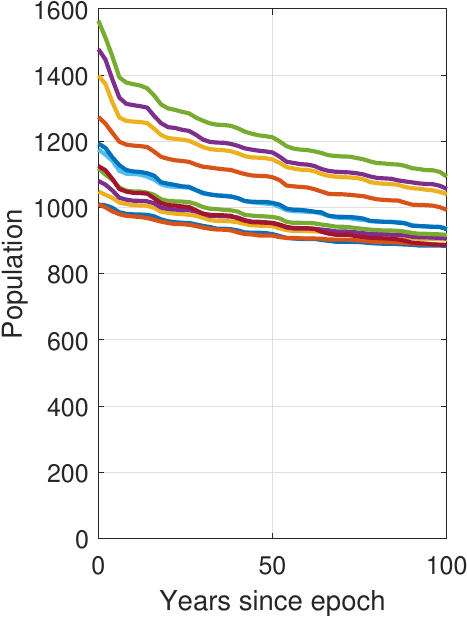}}		  
 
    \caption{Number of objects per object type for the No Future Launch cases with varying epochs}
    \label{fig:appendix-historicalNFL-SDNB}
    % Validation_results\0805_historicNFL\analysis_0805.m

\end{figure}

The \textit{No Future Launches} is run for a range of epochs, and Figure \ref{fig:appendix-historicalNFL-SDNB} shows the evolution of the population grouped into each each type of object.  Note that the debris class here denotes any type of debris. The transition from the failed PMD of the payload class into the derelict class is clearly seen, along with the increased debris population.  

\begin{table}[ht!]
    \centering
    % \small
    \caption{Breakdown of object type in the initial population for January 2023}
    \begin{tabular}{@{}ccccc@{}}
    \toprule
       Payload & Payload MRO & Payload Frag. Debris &  Payload Debris & Rocket Body \\ 
      7866 & 233 & 5573 & 95 & 971  \\    \midrule \midrule
      Rocket Body MRO  & Rocket Frag. Debris & Rocket Debris & Other Debris & Unknown  \\
      609 & 2808 & 25 & 253 & 1 \\
      \bottomrule
    \end{tabular}
    \label{tab:TLEinitpop}
    % djangSBX\AAS 2023\initialPopulationProfile.m
\end{table}

% \clearpage

\subsection{Megaconstellation future traffic model}

Table \ref{tab:megalaunch} describes the future megaconstellation modeled in this work compiled from various sources including \cite{mcdowell2023jonathan} as of March 2023.  \textit{Total num} column denotes the final operational constellation size without spares,  \textit{Start Year} and \textit{Finish Year} denotes the start and end of the ramp-up phase to reach the final operational number per constellation.  Much of the data for proposed megaconstellations are unavailable such as launch years and physical attributes, which are noted as dashes in the table.  The missing launch dates were estimated for missing dates.  The physical attributes were set equal to the Starlink constellation, as it is the only on-orbit megaconstellation yet, and most of these megaconstellations are also for communication services as well.

\begin{table}[!ht]
\small
\centering
\ra{1.3}
\caption{Modeled future traffic for megaconstellations}
\begin{tabular}{@{}lrrrcccc@{}} % llllllll
\toprule
Constellation & \begin{tabular}[c]{@{}c@{}}Alt \\ (km)\end{tabular} & \begin{tabular}[c]{@{}c@{}}Inc \\ (deg)\end{tabular} & \begin{tabular}[c]{@{}c@{}}Total \\ Num\end{tabular} & \begin{tabular}[c]{@{}c@{}}Start \\ Year\end{tabular} & \begin{tabular}[c]{@{}c@{}}Finish \\ Year\end{tabular} & \begin{tabular}[c]{@{}c@{}}Mass \\ (kg)\end{tabular} & \begin{tabular}[c]{@{}c@{}}Radius \\ (m)\end{tabular} \\ \midrule 
Starlink      & 550                                                 & 53                                                   & 1584                                                          & 2018  & 2027   & 260                                                  & 2.0                                                     \\
Starlink      & 570                                                 & 70                                                   & 720                                                           & 2018  & 2027   & 260                                                  & 2.0                                                     \\
Starlink      & 560                                                 & 97.6                                                 & 348                                                           & 2018  & 2027   & 260                                                  & 2.0                                                     \\
Starlink      & 540                                                 & 53.2                                                 & 1584                                                          & 2018  & 2027   & 260                                                  & 2.0                                                     \\
Starlink      & 560                                                 & 97.6                                                 & 172                                                           & 2018  & 2027   & 260                                                  & 2.0                                                     \\
Starlink2A    & 530                                                 & 43                                                   & 2500                                                          & 2023  & 2031   & 800                                                  & 2.0                                                     \\
Starlink2A    & 525                                                 & 53                                                   & 2500                                                          & 2023  & 2031   & 800                                                  & 2.0                                                     \\
Starlink2A    & 535                                                 & 33                                                   & 2500                                                          & 2023  & 2031   & 800                                                  & 2.0                                                     \\
Starlink2     & 340                                                 & 53                                                   & 5280                                                          & 2023  & 2031   & 1250                                                 & 4.0                                                     \\
Starlink2     & 345                                                 & 46                                                   & 5280                                                          & 2023  & 2031   & 1250                                                 & 4.0                                                     \\
Starlink2     & 350                                                 & 38                                                   & 5280                                                          & 2023  & 2031   & 1250                                                 & 4.0                                                     \\
Starlink2     & 360                                                 & 96.9                                                 & 3600                                                          & 2023  & 2031   & 1250                                                 & 4.0                                                     \\
Starlink2     & 530                                                 & 43                                                   & 860                                                           & 2023  & 2031   & 1250                                                 & 4.0                                                     \\
Starlink2     & 525                                                 & 53                                                   & 860                                                           & 2023  & 2031   & 1250                                                 & 4.0                                                     \\
Starlink2     & 535                                                 & 33                                                   & 860                                                           & 2023  & 2031   & 1250                                                 & 4.0                                                     \\
Starlink2     & 604                                                 & 148                                                  & 144                                                           & 2023  & 2031   & 1250                                                 & 4.0                                                     \\
Starlink2     & 614                                                 & 115.7                                                & 324                                                           & 2023  & 2031   & 1250                                                 & 4.0                                                     \\ \midrule
OneWeb        & 1200                                                & 87.9                                                 & 588                                                           & 2019  & 2023   & 150                                                  & 0.5                                                   \\
OneWeb        & 1200                                                & 55                                                   & 128                                                           & 2019  & 2023   & 150                                                  & 0.5                                                   \\
OneWeb        & 1200                                                & 87.9                                                 & 1764                                                          & 2025  & 2028   & 150                                                  & 0.5                                                   \\
OneWeb        & 1200                                                & 40                                                   & 2304                                                          & 2025  & 2028   & 150                                                  & 0.5                                                   \\
OneWeb        & 1200                                                & 55                                                   & 2304                                                          & 2025  & 2028   & 150                                                  & 0.5                                                   \\ \midrule
Kuiper        & 590                                                 & 33                                                   & 782                                                           & 2024  & 2029   & 700                                                  & 1.5                                                   \\
Kuiper        & 590                                                 & 30                                                   & 2                                                             & 2024  & 2029   & 700                                                  & 1.5                                                   \\
Kuiper        & 610                                                 & 42                                                   & 1292                                                          & 2024  & 2029   & 700                                                  & 1.5                                                   \\
Kuiper        & 630                                                 & 51.9                                                 & 1156                                                          & 2024  & 2029   & 700                                                  & 1.5       \\  \bottomrule
\end{tabular}
\label{tab:megalaunch}
\end{table}

\clearpage

\begin{table}[!ht]
\small
\centering
\ra{1.3}
\caption* {\textbf{Table \ref{tab:megalaunch} Continued:} Modeled future traffic for megaconstellations}
\begin{tabular}{@{}lrrrcccc@{}} 
\toprule
Constellation & \begin{tabular}[c]{@{}c@{}}Alt \\ (km)\end{tabular} & \begin{tabular}[c]{@{}c@{}}Inc \\ (deg)\end{tabular} & \begin{tabular}[c]{@{}c@{}}Total \\ Num\end{tabular} & \begin{tabular}[c]{@{}c@{}}Start \\ Year\end{tabular} & \begin{tabular}[c]{@{}c@{}}Finish \\ Year\end{tabular} & \begin{tabular}[c]{@{}c@{}}Mass \\ (kg)\end{tabular} & \begin{tabular}[c]{@{}c@{}}Radius \\ (m)\end{tabular} \\ \midrule 
Guanwang      & 590                                                 & 85.0                                                   & 480                                                           & 2035  & 2055   &   -                                                   &               -                                        \\
Guanwang      & 600                                                 & 50.0                                                   & 2000                                                          & 2035  & 2055   &   -                                                   &               -                                        \\
Guanwang      & 508                                                 & 60.0                                                 & 3600                                                          & 2035  & 2055   &   -                                                   &                -                                       \\
Guanwang      & 1145                                                & 30.0                                                   & 1728                                                          & 2035  & 2055   &  -                                                    &                -                                       \\
Guanwang      & 1145                                                & 40.0                                                   & 1728                                                          & 2035  & 2055   &  -                                                    &               -                                        \\
Guanwang      & 1145                                                & 50.0                                                   & 1728                                                          & 2035  & 2055   &   -                                                   &               -                                        \\
Guanwang      & 1145                                                & 60.0                                                   & 1728                                                          & 2035  & 2055   &  -                                                    &               -                                               \\ \midrule
Yinhe         & 511                                                 & 63.5                                                 & 1000                                                          & -  &   -     & 230                                                  & 0.7                                                   \\ \midrule
Hanwha        & 500                                                 & 97.5                                                 & 2000                                                          & 2025  & 2035   &   -                                                   &   -                                                    \\ \midrule
Lynk          & 500                                                 & 97.5                                                 & 2000                                                          & -  & -   & 125                                                  & 0.5                                                   \\ \midrule
Astra         & 700                                                 & 0                                                    & 40                                                            &  -     &  -      & 500                                                  &  -                                                     \\
Astra         & 690                                                 & 98.0                                                   & 504                                                           &  -     &  -      & 500                                                  &                -                                       \\
Astra         & 700                                                 & 55.0                                                   & 1792                                                          &  -     &   -     & 500                                                  &                -                                       \\
Astra         & 380                                                 & 97.0                                                   & 2240                                                          &  -     &  -      & 500                                                  &                -                                       \\
Astra         & 390                                                 & 30.0                                                   & 4896                                                          &  -     &  -      & 500                                                  &                -                                       \\
Astra         & 400                                                 & 55.0                                                   & 4148                                                          &  -     &  -      & 500                                                  &               -                                        \\ \midrule
Boeing        & 1056                                                & 54.0                                                   & 132                                                           & 2025  & 2030   &    -                                                  &                 -                                      \\ \midrule
Telesat       & 1015                                                & 99.0                                                   & 78                                                            & 2023  &   -     &   -                                                   &                -                                       \\
Telesat       & 1325                                                & 50.9                                                 & 220                                                           & 2023  &   -     &    -                                                  &               -                                        \\
Telesat       & 1015                                                & 99.0                                                   & 351                                                           & 2023  &   -     &   -                                                   &               -                                        \\
Telesat       & 1325                                                & 50.9                                                 & 1320                                                          & 2023  &   -     &    -                                                  &               -                                        \\ \midrule
HVNET         & 1150                                                & 55.0                                                   & 1440                                                          &   -    &   -     &    -                                                  &               -                                        \\ \midrule
SpinLaunch    & 830                                                 & 55.0                                                   & 1190                                                          &   -    &   -     & 150                                                  &               -                                        \\ \midrule
Globalstar3   & 485                                                 & 55.0                                                   & 1260                                                          &  -     &  -      &    -                                                  &               -                                        \\
Globalstar3   & 515                                                 & 70.0                                                   & 100                                                           &  -    &    -    &     -                                                 &               -                                        \\
Globalstar3   & 600                                                 & 55.0                                                   & 900                                                           &  -     &   -     &   -                                                   &               -                                        \\
Globalstar3   & 620                                                 & 98.0                                                   & 100                                                           &  -     &   -     &  -                                                   &                -                                       \\
Globalstar3   & 700                                                 & 55.0                                                   & 720                                                           &  -     &   -     &   -                                                   &              -                                        \\ \bottomrule
\end{tabular}
\label{tab:megalaunch2}
\end{table}
% formatting: https://people.inf.ethz.ch/markusp/teaching/guides/guide-tables.pdf

\begin{figure}[!ht]
    \centering    \includegraphics[width = 0.65\textwidth]{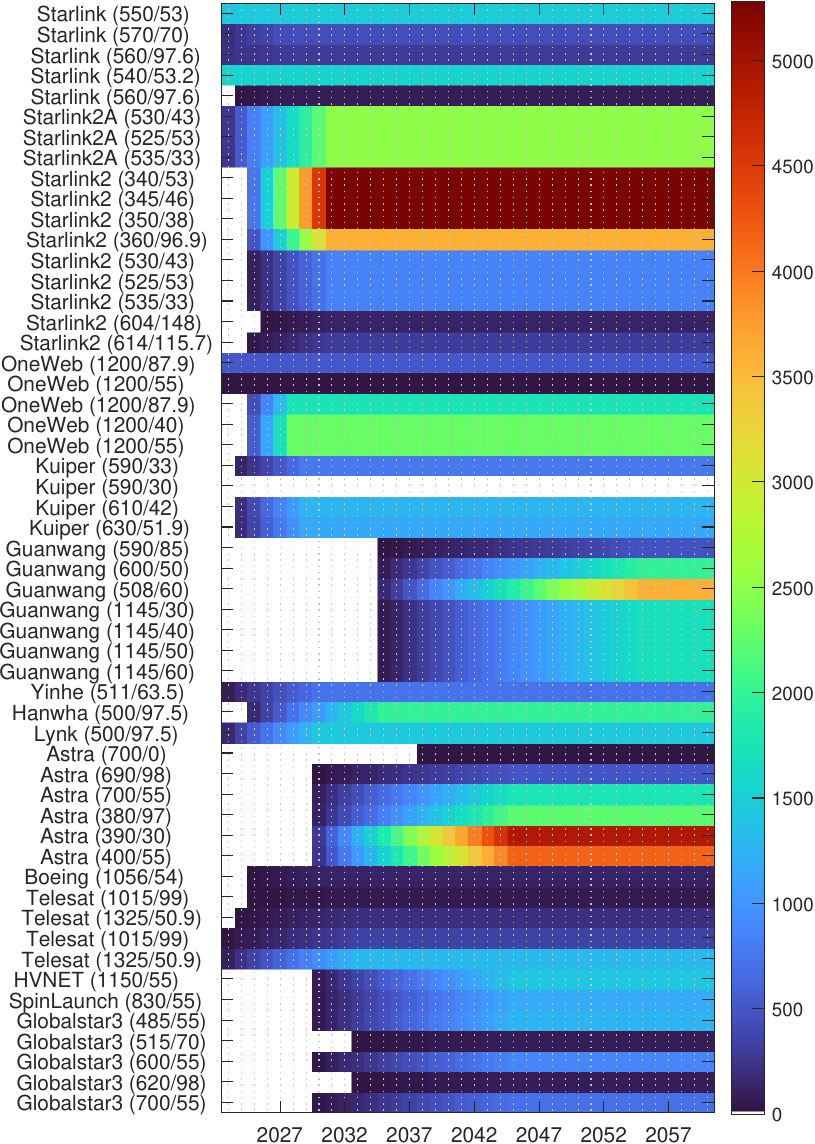}
    \caption{Megaconstellation population assumption}
    \label{fig:megalaunch-long}
\end{figure}

Figure \ref{fig:megalaunch-long} shows the unifed launch population per time broken into individual constellation planes.  The constellation name denotes the altitude in km and inclination in degrees in the parenthesis.  Note that \textit{Kuiper (590/30)} includes only two operational satellites.

\subsection{Megaconstellation future traffic model results}

\begin{figure}[!h]
    \centering
	\subcaptionbox{Starlink (v1, 1.5), Kuiper, OneWeb}[.35\textwidth]{\includegraphics[width=0.9\linewidth]{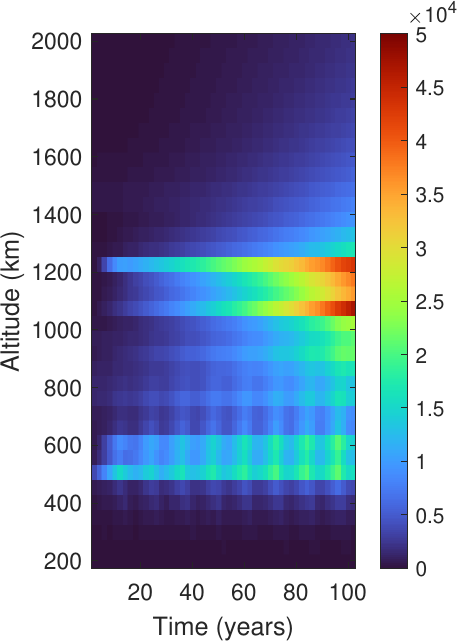}}  
    \subcaptionbox{Starlink (v1, 1.5, 2), Kuiper, OneWeb}[.35\textwidth]{\includegraphics[width=0.9\linewidth]{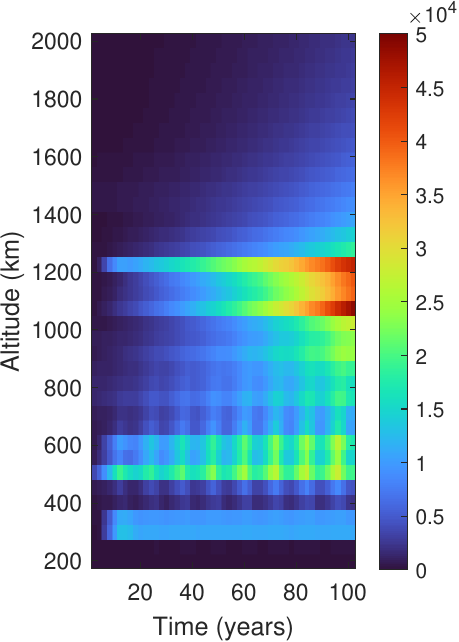}}
	\caption{Total population (>10 cm) per altitude for future launches}
	\label{fig:mega2-3}
\end{figure}

Other scenarios that compare portions of megaconstellation launches are shown in this section.  Figure \ref{fig:mega2-3} shows the evolution of the LEO environment with Starlink, Kuiper, and OneWeb launches, and compares the inclusion and exclusion of Starlink v2, the largest proposed megaconsetllation considered in this paper.  The plots are binned with an altitude bin of 50 km and a time bin of 1 year.  

\begin{figure}[!ht]
    \centering
	\subcaptionbox{$<600$ km megaconstellations}[.35\textwidth]{\includegraphics[width=0.9\linewidth]{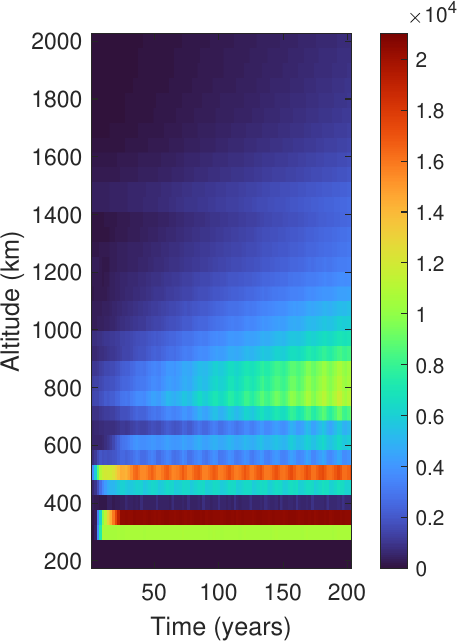}}
	\subcaptionbox{$<900$ km}[.35\textwidth]{\includegraphics[width=0.9\linewidth]{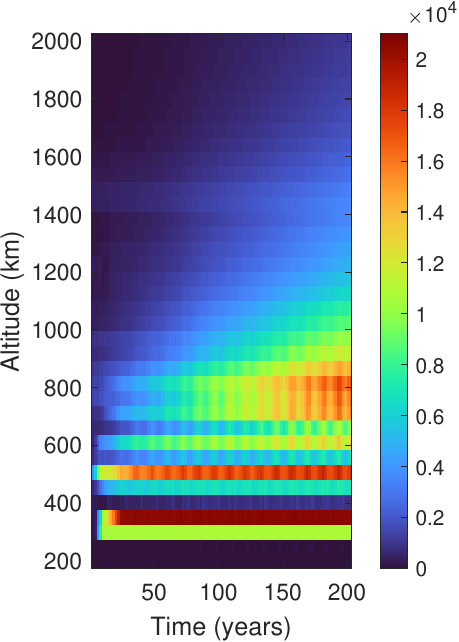}}  
	\caption{Population (>10 cm) evolution for megaconstellation launches limited to $<600$ km and to $<900$ km}
	\label{fig:megaAltLim600vs900}
    % 0730_megasweep/analysis.m
\end{figure}

Figure \ref{fig:megaAltLim600vs900} compares the evolution of megaconstellation launches that are limited to $<600$ km (total of 59336 operational megaconstellation payloads) and $<900$ km (total of 66598 operational payloads) constellations.  The plots are binned with an altitude bin of 50 km and a time bin of 1 year.

% \clearpage

\section*{Acknowledgments}
This research was sponsored by MIT Lincoln Scholar Fellowship and the United States Air Force Research Laboratory and the Department of the Air Force Artificial Intelligence Accelerator under Cooperative Agreement Number FA8750-19-2-1000. The views and conclusions contained in this document are those of the authors and should not be interpreted as representing the official policies, either expressed or implied, of the Department of the Air Force or the U.S. Government. The U.S. Government is authorized to reproduce and distribute reprints for Government purposes notwithstanding any copyright notation herein.  The authors acknowledge the MIT SuperCloud and Lincoln Laboratory Supercomputing Center for providing the HPC, database, and consultation resources that have contributed to the research results reported within this paper.

\bibliography{referencesNew}

\end{document}

%% file: nomenclature.tex
%% Acronyms
% for AIAA: these are automatically sorted as needed
\nomenclature[A]{}{}
\nomenclature[A]{AMR}{Area to Mass Ratio}
\nomenclature[A]{CARA}{Conjunction Assessment Risk Analysis}
\nomenclature[A]{ECI}{Earth Centered Inertial Coordinate System}
\nomenclature[A]{ESA}{European Space Agency}
\nomenclature[A]{FCC}{Federal Communications Commission}
\nomenclature[A]{GEO}{Geosynchronous Orbit}
\nomenclature[A]{ISS}{International Space Station}
\nomenclature[A]{ITU}{International Telecommunication Union}
\nomenclature[A]{LEO}{Low Earth Orbit (LEO)}
\nomenclature[A]{LNT}{Lethal Non-Trackable}
\nomenclature[A]{MASTER}{ESA's Meteoroid And Space debris Terrestrial Environment Reference}
\nomenclature[A]{MC}{Monte Carlo}
\nomenclature[A]{MOCAT}{MIT Orbital Capacity Analysis Tool}
% \nomenclature[A]{NASA}{National Aeronautics and Space Administration}
% \nomenclature[A]{RSO}{Resident Space Objects}
\nomenclature[A]{SGP4}{Simplified General Perturbations Satellite Orbit Model 4}
\nomenclature[A]{SSEM}{Source Sink Evolutionary Model}
\nomenclature[A]{TLE}{Two-Line Element}
\nomenclature[A]{CAM}{Collision Avoidance Maneuver}
\nomenclature[A]{PL}{Payload}
\nomenclature[A]{MRO}{Mission Related Objects}
\nomenclature[A]{FD}{Fragmentation Debris}
\nomenclature[A]{RB}{Rocket Body}
\nomenclature[A]{MITRI}{MIT Risk Index}
\nomenclature[A]{NFL}{No Future Launch}

\nomenclature[B]{$A$}{Area}
\nomenclature[B]{$a$}{Semi-major axis}
\nomenclature[B]{$f$}{True anomaly}
\nomenclature[B]{$C_D$}{Drag coefficient}
\nomenclature[B]{$e$}{Orbit eccentricity}
\nomenclature[B]{$i$}{Inclination}
\nomenclature[B]{$J_2$}{Earth's oblateness}
\nomenclature[B]{$L_C$}{Characteristic length}
\nomenclature[B]{$m$}{Mass}
\nomenclature[B]{$P_C$}{Probability of collision}
\nomenclature[B]{$PDF$}{Probability density function}
\nomenclature[B]{$t$}{Time}
\nomenclature[B]{$\Omega$}{Right Ascension of the Ascending Node (RAAN)}
\nomenclature[B]{$\omega$}{Argument of perigee}
\nomenclature[B]{$\sigma$}{Cross sectional area}
\nomenclature[B]{$M$}{Mean anomaly}
\nomenclature[B]{$R_E$}{Earth radius}
\nomenclature[B]{$\rho$}{Atmospheric density}
\nomenclature[B]{$B^*$}{$B^*$ or Bstar parameter}
\nomenclature[B]{$r$}{Radius of object}
\nomenclature[B]{$r_{x,y,z}$}{ECI position vector components}
\nomenclature[B]{$v_{x,y,z}$}{ECI velocity vector components}
\nomenclature[B]{$v_c$}{Collision velocity}
\nomenclature[B]{T}{Temperature}
\nomenclature[B]{$A_p$}{Geomagnetic planetary index}
\nomenclature[B]{$s_i$}{Spatial density of objects of type $i$}
\nomenclature[B]{$dU$}{Inverse of the volume of cube}
\nomenclature[B]{$\Tilde{E}_p $}{Specific energy in a collision}
\nomenclature[B]{$\alpha$}{Collision avoidance failure rate between active and non-active objects}
\nomenclature[B]{$\alpha_a$}{Collision avoidance failure rate between active objects}
\nomenclature[B]{$P_{PMD}$}{Probability of succesful PMD}
\nomenclature[B]{$\Delta t$}{Simulation time step}
\nomenclature[B]{CDM}{Conjunction data message}
\printnomenclature

%% file: main_test.bbl
\begin{thebibliography}{100}
\newcommand{\enquote}[1]{``#1''}
\providecommand{\natexlab}[1]{#1}
\providecommand{\url}[1]{\texttt{#1}}
\providecommand{\urlprefix}{URL }
\expandafter\ifx\csname urlstyle\endcsname\relax
  \providecommand{\doi}[1]{\discretionary{}{}{}https://doi.org/#1}\else
  \providecommand{\doi}[1]{\discretionary{}{}{}\urlstyle{rm}\url{https://doi.org/#1}}\fi

\bibitem[{SpaceX(2022)}]{SpaceX2022}
SpaceX, \enquote{Updates,} , 2 2022.
\newblock \urlprefix\url{https://www.spacex.com/updates/index.html#sustainability}.

\bibitem[{Kessler and Cour-Palais(1978)}]{Kessler1978}
Kessler, D.~J., and Cour-Palais, B.~G., \enquote{Collision Frequency of Artificial Satellites: The Creation of a Debris Belt.} \emph{J Geophys Res}, Vol.~83, 1978, pp. 2637--2646.
\newblock \doi{10.1029/JA083iA06p02637}.

\bibitem[{Kessler et~al.(2010)Kessler, Johnson, Liou, and Matney}]{Kessler2010}
Kessler, D.~J., Johnson, N.~L., Liou, J.-C., and Matney, M., \enquote{The Kessler Syndrome: Implications to Future Space operations,} \emph{Advances in the Astronautical Sciences}, Vol. 137, 2010, pp. 47--61.

\bibitem[{Liou(2011)}]{Liou2011}
Liou, J.~C., \enquote{An Active Debris Removal Parametric Study for LEO Environment Remediation,} \emph{Advances in Space Research}, Vol.~47, 2011, pp. 1865--1876.
\newblock \doi{10.1016/J.ASR.2011.02.003}.

\bibitem[{White and Lewis(2014{\natexlab{a}})}]{White2014AnRemoval}
White, A.~E., and Lewis, H.~G., \enquote{{An adaptive strategy for active debris removal},} \emph{Advances in Space Research}, Vol.~53, No.~8, 2014{\natexlab{a}}, pp. 1195--1206.
\newblock \doi{10.1016/j.asr.2014.01.021}.

\bibitem[{Hakima and Emami(2018)}]{Hakima2018AssessmentDebris}
Hakima, H., and Emami, M.~R., \enquote{{Assessment of active methods for removal of LEO debris},} \emph{Acta Astronautica}, Vol. 144, 2018, pp. 225--243.
\newblock \doi{10.1016/j.actaastro.2017.12.036}.

\bibitem[{FCC(2022{\natexlab{a}})}]{FCCdebris2022}
\enquote{Space Innovation: Mitigation of Orbital Debris in the New Space Age [Press Release],} , 9 2022{\natexlab{a}}.
\newblock \urlprefix\url{www.fcc.gov}.

\bibitem[{Johnson(2017)}]{SWFhandbookForNewActors2017}
Johnson, C.~D., \enquote{Handbook For New Actors In Space,} , 2017.

\bibitem[{{Space Safety Coalition}(2019)}]{SSCbestPractices2019}
{Space Safety Coalition}, \enquote{Best Practices for the Sustainability of Space Operations,} , 9 2019.

\bibitem[{{SpaceX, OneWeb, and Iridium}(2022)}]{AIAAbestPractices2022}
{SpaceX, OneWeb, and Iridium}, \enquote{Satellite Orbital Safety Best Practices,} , 9 2022.

\bibitem[{{Executive Office of the President}(2018)}]{SPD3_2018}
{Executive Office of the President}, \enquote{National Space Traffic Management Policy,} \emph{Presidential Document}, Vol. 2018-13521, 2018, pp. 28969--28976.
\newblock \urlprefix\url{https://www.federalregister.gov/documents/2018/06/21/2018-13521/national-space-traffic-management-policy}.

\bibitem[{{ESA Space Debris Office}(2022)}]{ESAspaceEnvReport2022}
{ESA Space Debris Office}, \enquote{{ESA} Annual Space Environment Report,} , 4 2022.

\bibitem[{Colvin et~al.(2023)Colvin, Karcz, and Wusk}]{colvin2023cost}
Colvin, T.~J., Karcz, J., and Wusk, G., \enquote{Cost and benefit analysis of orbital debris remediation,} \emph{NASA Office of Technology, Policy, and Strategy}, 2023.
\newblock \urlprefix\url{https://ntrs.nasa.gov/citations/20230002817}.

\bibitem[{FCC(2022{\natexlab{b}})}]{FCC5yrRule}
FCC, \enquote{Space Innovation {IB} Docket No. 22-271 Mitigation of Orbital Debris in the New Space Age,} Docket No: 22-271, 18-313, 12 2022{\natexlab{b}}.
\newblock \urlprefix\url{https://docs.fcc.gov/public/attachments/FCC-22-74A1_Rcd.pdf}.

\bibitem[{{United Nations}(2022)}]{antiASAT2022UN}
{United Nations}, \enquote{Destructive direct-ascent anti-satellite missile testing : resolution / adopted by the General Assembly,} , 12 2022.
\newblock \urlprefix\url{https://digitallibrary.un.org/record/3996915}.

\bibitem[{ant(2024)}]{antiASAT2024SWF}
\enquote{Space Industry Statement in Support of International Commitments Not To Conduct Destructive Anti-Satellite Testing,} , 3 2024.
\newblock \urlprefix\url{https://swfound.org/media/207814/da-asat-statement_march-2024.pdf}.

\bibitem[{Letizia et~al.(2017)Letizia, Colombo, Lewis, and Krag}]{Letizia2017}
Letizia, F., Colombo, C., Lewis, H., and Krag, H., \enquote{Extending the {ECOB} space debris index with fragmentation risk estimation,} \emph{7th European Conference on Space Debris}, 2017.

\bibitem[{Krag et~al.(2018)Krag, Lemmens, and Letizia}]{kragCapacity}
Krag, H., Lemmens, S., and Letizia, F., \enquote{Space traffic management through the control of the space environment’s capacity,} \emph{5th European Workshop on Space Debris Modeling and Remediation}, 2018.

\bibitem[{Servadio et~al.(0)Servadio, Simha, Gusmini, Jang, St.~Francis, D’Ambrosio, Lavezzi, and Linares}]{MITRI}
Servadio, S., Simha, N., Gusmini, D., Jang, D., St.~Francis, T., D’Ambrosio, A., Lavezzi, G., and Linares, R., \enquote{Risk Index for the Optimal Ranking of Active Debris Removal Targets,} \emph{Journal of Spacecraft and Rockets}, Vol.~0, No.~0, 0, pp. 1--14.
\newblock \doi{10.2514/1.A35752}.

\bibitem[{Rossi et~al.(2015)Rossi, Valsecchi, and Alessi}]{Rossi2015}
Rossi, A., Valsecchi, G.~B., and Alessi, E.~M., \enquote{The Criticality of Spacecraft Index,} \emph{Advances in Space Research}, Vol.~56, 2015, pp. 449--460.
\newblock \doi{10.1016/J.ASR.2015.02.027}.

\bibitem[{Letizia et~al.(2016)Letizia, Colombo, Lewis, and Krag}]{Letizia2016severity}
Letizia, F., Colombo, C., Lewis, H.~G., and Krag, H., \enquote{Assessment of breakup severity on operational satellites,} \emph{Advances in Space Research}, Vol.~58, 2016, pp. 1255--1274.
\newblock \doi{10.1016/J.ASR.2016.05.036}.

\bibitem[{McKnight et~al.(2021)McKnight, Witner, Letizia, Lemmens, Anselmo, Pardini, Rossi, Kunstadter, Kawamoto, Aslanov, Perez, Ruch, Lewis, Nicolls, Jing, Dan, Dongfang, Baranov, and Grishko}]{mcknight2021}
McKnight, D., Witner, R., Letizia, F., Lemmens, S., Anselmo, L., Pardini, C., Rossi, A., Kunstadter, C., Kawamoto, S., Aslanov, V., Perez, J. C.~D., Ruch, V., Lewis, H., Nicolls, M., Jing, L., Dan, S., Dongfang, W., Baranov, A., and Grishko, D., \enquote{Identifying the 50 statistically-most-concerning derelict objects in {LEO},} \emph{Acta Astronautica}, Vol. 181, 2021, pp. 282--291.
\newblock \doi{10.1016/j.actaastro.2021.01.021}.

\bibitem[{Volterra(1926)}]{volterra1926fluctuations}
Volterra, V., \enquote{Fluctuations in the abundance of a species considered mathematically,} \emph{Nature}, Vol. 118, No. 2972, 1926, pp. 558--560.

\bibitem[{Lotka(1920)}]{lotka1920analytical}
Lotka, A.~J., \enquote{Analytical note on certain rhythmic relations in organic systems,} \emph{Proceedings of the National Academy of Sciences}, Vol.~6, No.~7, 1920, pp. 410--415.
\newblock \doi{10.1073/pnas.6.7.410}.

\bibitem[{Talent(1992)}]{talent1990pib}
Talent, D.~L., \enquote{Analytic model for orbital debris environmental management,} \emph{Journal of Spacecraft and Rockets}, Vol.~29, No.~4, 1992, pp. 508--513.
\newblock \doi{10.2514/3.25493}.

\bibitem[{Lewis et~al.(2009{\natexlab{a}})Lewis, Swinerd, Newland, and Saunders}]{Lewis2009}
Lewis, H.~G., Swinerd, G.~G., Newland, R.~J., and Saunders, A., \enquote{The fast debris evolution model,} \emph{Advances in Space Research}, Vol.~44, 2009{\natexlab{a}}, pp. 568--578.
\newblock \doi{10.1016/j.asr.2009.05.018}.

\bibitem[{{Long, G.}(2020)}]{JASONreport2020TheSatellites}
{Long, G.}, \enquote{{The Impacts of Large Constellations of Satellites},} \emph{The MITRE Corporation, JASON}, 2020.

\bibitem[{Lifson et~al.(2023)Lifson, Jang, Pasiecznik, and Linares}]{MOCATSSEM2023}
Lifson, M., Jang, D., Pasiecznik, C., and Linares, R., \enquote{{MOCAT-SSEM}: A Source-Sink Evolutionary Model for Space Debris Environment Evolutionary Modeling,} \emph{9th Annual Space Traffic Management Conference}, 2023.

\bibitem[{D’Ambrosio and Linares(2024)}]{d2024carrying}
D’Ambrosio, A., and Linares, R., \enquote{Carrying Capacity of Low Earth Orbit Computed Using Source-Sink Models,} \emph{Journal of Spacecraft and Rockets}, 2024, pp. 1--17.
\newblock \doi{10.2514/1.A35729}.

\bibitem[{Gusmini et~al.(0)Gusmini, D’Ambrosio, Servadio, Siew, Di~Lizia, and Linares}]{gusmini2024effects}
Gusmini, D., D’Ambrosio, A., Servadio, S., Siew, P.~M., Di~Lizia, P., and Linares, R., \enquote{Effects of Orbit Raising and Deorbiting in Source-Sink Evolutionary Models,} \emph{Journal of Spacecraft and Rockets}, Vol.~0, No.~0, 0, pp. 1--14.
\newblock \doi{10.2514/1.A35849}.

\bibitem[{D’Ambrosio et~al.(2023)D’Ambrosio, Servadio, Mun~Siew, and Linares}]{d2023novel}
D’Ambrosio, A., Servadio, S., Mun~Siew, P., and Linares, R., \enquote{Novel Source–Sink Model for Space Environment Evolution with Orbit Capacity Assessment,} \emph{Journal of Spacecraft and Rockets}, Vol.~60, No.~4, 2023, pp. 1112--1126.
\newblock \doi{10.2514/1.A35579}.

\bibitem[{Krisko(2014)}]{Krisko2014TheInvited}
Krisko, P.~H., \enquote{{The New NASA Orbital Debris Engineering Model ORDEM 3.0 (Invited)},} \emph{AIAA/AAS Astrodynamics Specialist Conference}, 2014.
\newblock \doi{10.2514/6.2014-4227}.

\bibitem[{Liou et~al.(2004)Liou, Hall, Krisko, and Opiela}]{LEGEND2004}
Liou, J.~C., Hall, D.~T., Krisko, P.~H., and Opiela, J.~N., \enquote{{LEGEND} – a Three-dimensional {LEO-to-GEO} Debris Evolutionary Model,} \emph{Advances in Space Research}, Vol.~34, 2004, pp. 981--986.
\newblock \doi{10.1016/J.ASR.2003.02.027}.

\bibitem[{Lewis et~al.(2001)Lewis, Swinerd, Williams, and Gittins}]{Lewis2001}
Lewis, H., Swinerd, G., Williams, N., and Gittins, G., \enquote{DAMAGE: a Dedicated GEO Debris Model Framework,} \emph{3rd European Conference on Space Debris}, 2001, pp. 373--378.

\bibitem[{Lewis et~al.(2011)Lewis, Saunders, Swinerd, and Newland}]{Lewis2011}
Lewis, H.~G., Saunders, A., Swinerd, G., and Newland, R.~J., \enquote{Effect of thermospheric contraction on remediation of the near-Earth space debris environment,} \emph{Journal of Geophysical Research: Space Physics}, Vol. 116, 2011.
\newblock \doi{10.1029/2011JA016482}.

\bibitem[{Wang and Liu(2019)}]{Wang2019AnSOLEM}
Wang, X.~W., and Liu, J., \enquote{An Introduction to a New Space Debris Evolution Model: SOLEM,} \emph{Advances in Astronomy}, Vol. 2019, 2019.
\newblock \doi{10.1155/2019/2738276}.

\bibitem[{Dolado-Perez et~al.(2013)Dolado-Perez, Di~Costanzo, and Revelin}]{medee}
Dolado-Perez, J.-C., Di~Costanzo, R., and Revelin, B., \enquote{{Introducing {MEDEE} - a New Orbital Debris Evolutionary Model},} \emph{Proc. 6th European Conference on Space Debris}, 2013, pp. 22--25.

\bibitem[{Martin et~al.(2004)Martin, Cheese, and Klinkrad}]{DELTA2}
Martin, C.~E., Cheese, J.~E., and Klinkrad, H., \enquote{{Space Debris Environment Analysis with DELTA 2.0},} \emph{International Astronautical Federation - 55th International Astronautical Congress}, 2004.
\newblock \doi{10.2514/6.IAC-04-IAA.5.12.5.04}.

\bibitem[{Radtke et~al.(2017)Radtke, Mueller, Schaus, and Stoll}]{luca2}
Radtke, J., Mueller, S., Schaus, V., and Stoll, E., \enquote{{LUCA2 - An enhanced long-term utility for collision analysis},} \emph{Proc. 7th European Conference on Space Debris}, 2017.

\bibitem[{KAWAMOTO et~al.(2018)KAWAMOTO, HIRAI, KITAJIMA, ABE, and HANADA}]{KAWAMOTO2018}
KAWAMOTO, S., HIRAI, T., KITAJIMA, S., ABE, S., and HANADA, T., \enquote{Evaluation of Space Debris Mitigation Measures Using a Debris Evolutionary Model,} \emph{Transactions of the Japan Society for Aeronautical and Space Sciences, Aerospace Technology Japan}, Vol.~16, 2018, pp. 599--603.
\newblock \doi{10.2322/TASTJ.16.599}.

\bibitem[{Drmola and Hubik(2018)}]{Drmola2018KesslerModel}
Drmola, J., and Hubik, T., \enquote{Kessler Syndrome: System Dynamics Model,} \emph{Space Policy}, Vol. 44-45, 2018, pp. 29--39.
\newblock \doi{10.1016/J.SPACEPOL.2018.03.003}.

\bibitem[{Rosengren et~al.(2019)Rosengren, Skoulidou, Tsiganis, and Voyatzis}]{Rosengren2019DynamicalOrbits}
Rosengren, A.~J., Skoulidou, D.~K., Tsiganis, K., and Voyatzis, G., \enquote{Dynamical cartography of Earth satellite orbits,} \emph{Advances in Space Research}, Vol.~63, 2019, pp. 443--460.
\newblock \doi{10.1016/J.ASR.2018.09.004}.

\bibitem[{Giudici et~al.(2024)Giudici, Colombo, Horstmann, Letizia, and Lemmens}]{Giudici2024}
Giudici, L., Colombo, C., Horstmann, A., Letizia, F., and Lemmens, S., \enquote{Density-based evolutionary model of the space debris environment in low-Earth orbit,} \emph{Acta Astronautica}, Vol. 219, 2024, pp. 115--127.
\newblock \doi{10.1016/j.actaastro.2024.03.008}.

\bibitem[{Johnson et~al.(2001)Johnson, Krisko, Liou, and Anz-Meador}]{nasaSTMevolve4}
Johnson, N.~L., Krisko, P.~H., Liou, J.~C., and Anz-Meador, P.~D., \enquote{{NASA}'s New Breakup Model of EVOLVE 4.0,} \emph{Advances in Space Research}, Vol.~28, 2001, pp. 1377--1384.
\newblock \doi{10.1016/S0273-1177(01)00423-9}.

\bibitem[{Liou et~al.(2003)Liou, Kessler, Matney, and Stansbery}]{LiouCUBE2003}
Liou, J.~C., Kessler, D.~J., Matney, M., and Stansbery, G., \enquote{A New Approach To Evaluate Collision Probabilities Among Asteroids, Comets, And Kuiper Belt Objects,} \emph{Lunar and Planetary Science Conference}, 2003.

\bibitem[{Liou(2006)}]{LiouCUBE2006}
Liou, J.~C., \enquote{Collision activities in the future orbital debris environment,} \emph{Advances in Space Research}, Vol.~38, 2006, pp. 2102--2106.
\newblock \doi{10.1016/j.asr.2005.06.021}.

\bibitem[{{MIT ARCLab}(2024)}]{MOCATMCgithub}
{MIT ARCLab}, \enquote{MOCAT-MC GitHub Repository,} \url{https://github.com/ARCLab-MIT/MOCAT-MC}, 2024.

\bibitem[{Vallado et~al.(2006)Vallado, Crawford, Hujsak, and Kelso}]{valladoSGP42006}
Vallado, D., Crawford, P., Hujsak, R., and Kelso, T., \enquote{Revisiting Spacetrack Report \#3,} \emph{AIAA/AAS Astrodynamics Specialist Conference and Exhibit}, 2006.
\newblock \doi{10.2514/6.2006-6753}.

\bibitem[{Martinusi et~al.(2015)Martinusi, Dell’Elce, and Kerschen}]{Martinusi2015}
Martinusi, V., Dell’Elce, L., and Kerschen, G., \enquote{Analytic propagation of near-circular satellite orbits in the atmosphere of an oblate planet,} \emph{Celestial Mechanics and Dynamical Astronomy}, Vol. 123, 2015, pp. 85--103.
\newblock \doi{10.1007/s10569-015-9630-7}.

\bibitem[{Cefola et~al.(2014)Cefola, Folcik, Di-Costanzo, Bernard, Setty, and Juan}]{cefola2014revisiting}
Cefola, P., Folcik, Z., Di-Costanzo, R., Bernard, N., Setty, S., and Juan, J., \enquote{Revisiting the DSST standalone orbit propagator,} \emph{Advances in the Astronautical Sciences}, Vol. 152, 2014, pp. 2891--2914.

\bibitem[{Vallado(2022)}]{Vallado2022}
Vallado, D.~A., \emph{Fundamentals of Astrodynamics and Applications}, 5\textsuperscript{th} ed., Microcosm Press, 2022.

\bibitem[{Horstmann et~al.(2020)Horstmann, Hesselbach, Wiedemann, Flegel, Oswald, and Krag}]{MASTER8}
Horstmann, A., Hesselbach, S., Wiedemann, C., Flegel, S., Oswald, M., and Krag, H., \enquote{Enhancement of {S/C} Fragmentation and Environment Evolution Models,} \emph{European Space Agency}, 2020.

\bibitem[{Bowman et~al.(2008)Bowman, Tobiska, Marcos, Huang, Lin, and Burke}]{bowman2008new}
Bowman, B., Tobiska, W.~K., Marcos, F., Huang, C., Lin, C., and Burke, W., \enquote{A new empirical thermospheric density model JB2008 using new solar and geomagnetic indices,} \emph{AIAA/AAS Astrodynamics Specialist Conference and Exhibit}, 2008, p. 6438.
\newblock \doi{https://doi.org/10.2514/6.2008-6438}.

\bibitem[{Tobiska et~al.(2008)Tobiska, Bowman, and Bouwer}]{tobiska2008solar}
Tobiska, W.~K., Bowman, B., and Bouwer, S.~D., \enquote{Solar and geomagnetic indices for the JB2008 thermosphere density model,} \emph{Contract}, Vol. 19628, No. 03-C, 2008, p. 0076.

\bibitem[{Klinkrad(1991)}]{DISCOS}
Klinkrad, H., \enquote{{DISCOS - ESA's database and information system characterising objects in space},} \emph{Advances in Space Research}, Vol.~11, No.~12, 1991, pp. 43--52.
\newblock \doi{10.1016/0273-1177(91)90541-Q}.

\bibitem[{Hoots and Roehrich(1980)}]{Hoots80}
Hoots, F.~R., and Roehrich, R.~L., \enquote{Spacetrack Report No. 3: Models for propagation of {NORAD} element sets,} Tech. rep., Aerospace Defense Center, Peterson Air Force Base, 1980.

\bibitem[{Acedo(2017)}]{acedo2017kinematics}
Acedo, L., \enquote{Kinematics effects of atmospheric friction in spacecraft flybys,} \emph{Advances in Space Research}, Vol.~59, No.~7, 2017, pp. 1715--1723.
\newblock \doi{https://doi.org/10.1016/j.asr.2017.01.008}.

\bibitem[{Lewis et~al.(2019)Lewis, Diserens, Maclay, and Sheehan}]{LewisCUBElimitations2019}
Lewis, H.~G., Diserens, S., Maclay, T., and Sheehan, J.~P., \enquote{Limitations of the cube method for assessing large constellations,} \emph{First International Orbital Debris Conference}, 2019.

\bibitem[{{Alarc{\'o}n-Rodr{\'\i}guez} et~al.(2002){Alarc{\'o}n-Rodr{\'\i}guez}, {Mart{\'\i}nez Fadrique}, and {Klinkrad}}]{Rodriguez2002}
{Alarc{\'o}n-Rodr{\'\i}guez}, J.~R., {Mart{\'\i}nez Fadrique}, F., and {Klinkrad}, H., \enquote{{Collision Risk Assessment with a `Smart Sieve' Method},} \emph{Joint ESA-NASA Space-Flight Safety Conference}, Vol. 486, 2002, p. 159.

\bibitem[{George(2011)}]{George2011}
George, E.~R., \enquote{A High Performance Conjunction Analysis Technique for Cluster and Multi-Core Computers,} \emph{Advanced Maui Optical and Space Surveillance Technologies (AMOS) Conference}, 2011.

\bibitem[{Lue(2011)}]{Lue2011}
Lue, A., \enquote{The All-Versus-All Low Earth Orbit Conjunction Problem,} \emph{Advanced Maui Optical and Space Surveillance Technologies (AMOS) Conference}, 2011.

\bibitem[{Frey and Colombo(2021)}]{Frey2021}
Frey, S., and Colombo, C., \enquote{Transformation of satellite breakup distribution for probabilistic orbital collision hazard analysis,} \emph{Journal of Guidance, Control, and Dynamics}, Vol.~44, 2021, pp. 88--105.
\newblock \doi{10.2514/1.G004939}.

\bibitem[{Finkleman et~al.(2008)Finkleman, Oltrogge, Faulds, and Gerber}]{Finkleman2008AnalysisEvents}
Finkleman, D., Oltrogge, D.~L., Faulds, A., and Gerber, J., \enquote{{Analysis of the response of a space surveillance network to orbital debris events},} \emph{AAS/AIAA Astrodynamics Specialist Conference}, 2008.

\bibitem[{Heard(1976)}]{Heard1976DispersionParticles}
Heard, W.~B., \enquote{{Dispersion of ensembles of non-interacting particles},} \emph{Astrophysics and Space Science}, Vol.~43, No.~1, 1976, pp. 63--82.
\newblock \doi{10.1007/BF00640556}.

\bibitem[{Cordelli(1991)}]{Cordelli1991TheModel}
Cordelli, A., \enquote{{The Proliferation of Orbiting Fragments: A Simple Mathematical Model},} \emph{Science {\&} Global Security}, Vol.~2, No.~4, 1991, pp. 365--378.
\newblock \doi{10.1080/08929889108426373}.

\bibitem[{Frazzoli et~al.(1996)Frazzoli, Palmerini, and Graziani}]{Frazzoli1996DebrisDesign}
Frazzoli, E., Palmerini, G.~B., and Graziani, F., \enquote{{Debris cloud evolution: Mathematical modelling and application to satellite constellation design},} \emph{Acta Astronautica}, Vol.~39, No.~6, 1996, pp. 439--445.
\newblock \doi{10.1016/S0094-5765(96)00156-7}.

\bibitem[{Rossi et~al.(2016)Rossi, Lewis, White, Anselmo, Pardini, Krag, and Bastida~Virgili}]{Rossi2016AnalysisOrbits}
Rossi, A., Lewis, H., White, A., Anselmo, L., Pardini, C., Krag, H., and Bastida~Virgili, B., \enquote{{Analysis of the consequences of fragmentations in low and geostationary orbits},} \emph{Advances in Space Research}, Vol.~57, No.~8, 2016, pp. 1652--1663.
\newblock \doi{10.1016/J.ASR.2015.05.035}.

\bibitem[{Lewis et~al.(2017)Lewis, Radtke, Rossi, Beck, Oswald, Anderson, Bastida~Virgili, and Krag}]{LewisSensitivity}
Lewis, H.~G., Radtke, J., Rossi, A., Beck, J., Oswald, M., Anderson, P., Bastida~Virgili, B., and Krag, H., \enquote{{Sensitivity of the Space Debris Environment to Large Constellations and Small Satellites},} \emph{Journal of the British Interplanetary Society}, Vol.~70, 2017, pp. 105--117.

\bibitem[{Letizia(2018)}]{Letizia2018ExtensionPropagation}
Letizia, F., \enquote{{Extension of the density approach for debris cloud propagation},} \emph{Journal of Guidance, Control, and Dynamics}, Vol.~41, No.~12, 2018, pp. 2650--2656.
\newblock \doi{10.2514/1.G003675}.

\bibitem[{Su and Kessler(1985)}]{Su1985ContributionEnvironment}
Su, S.~Y., and Kessler, D.~J., \enquote{{Contribution of explosion and future collision fragments to the orbital debris environment},} \emph{Advances in Space Research}, Vol.~5, No.~2, 1985, pp. 25--34.
\newblock \doi{10.1016/0273-1177(85)90384-9}.

\bibitem[{Ruch et~al.(2021)Ruch, Serra, Omaly, and Dolado~Perez}]{Ruch2021DECOUPLEDENVIRONMENT}
Ruch, V., Serra, R., Omaly, P., and Dolado~Perez, J.~C., \enquote{{Decoupled Analysis of the Effect of Past and Future Space Activity on the Orbital Environment},} \emph{8th European Conference on Space Debris}, 2021, pp. 20--23.

\bibitem[{Dell'Elce et~al.(2015)Dell'Elce, Arnst, and Kerschen}]{DellElce2015ProbabilisticCharacterization}
Dell'Elce, L., Arnst, M., and Kerschen, G., \enquote{{Probabilistic assessment of the lifetime of low-earth-orbit spacecraft: Uncertainty characterization},} \emph{Journal of Guidance, Control, and Dynamics}, Vol.~38, No.~5, 2015, pp. 900--912.
\newblock \doi{10.2514/1.G000148}.

\bibitem[{Luo and Yang(2017)}]{Luo2017AMechanics}
Luo, Y.~z., and Yang, Z., \enquote{{A review of uncertainty propagation in orbital mechanics},} \emph{Progress in Aerospace Sciences}, Vol.~89, 2017, pp. 23--39.
\newblock \doi{10.1016/J.PAEROSCI.2016.12.002}.

\bibitem[{Lewis(2020)}]{LewisUnderstandingDynamics}
Lewis, H.~G., \enquote{Understanding long-term orbital debris population dynamics,} \emph{Journal of Space Safety Engineering}, Vol.~7, No.~3, 2020, pp. 164--170.
\newblock \doi{https://doi.org/10.1016/j.jsse.2020.06.006}.

\bibitem[{Weeden(2013)}]{Weeden2013Anti-satelliteChina}
Weeden, B., \enquote{{Anti-satellite Tests in Space— The Case of China},} \emph{Secure World Foundation}, 2013.
\newblock \urlprefix\url{https://swfound.org/media/115643/china_asat_testing_fact_sheet_aug_2013.pdf}.

\bibitem[{Kelso(2009)}]{Kelso2009AnalysisCollision}
Kelso, T.~S., \enquote{{Analysis of the Iridium 33-Cosmos 2251 Collision},} \emph{19th AIAA/AAS Astrodynamics Specialist Conference}, 2009.
\newblock \urlprefix\url{http://celestrak.com/SOCRATES/}.

\bibitem[{Stansbery et~al.(2008)Stansbery, Matney, Liou, and Whitlock}]{Stansbery2008}
Stansbery, G., Matney, M., Liou, J.~C., and Whitlock, D., \enquote{A Comparison of Catastrophic On-Orbit Collisions,} \emph{Advanced Maui Optical and Space Surveillance Technologies (AMOS) Conference}, 2008.

\bibitem[{Braun et~al.(2017)Braun, Lemmens, Reihs, Krag, and Horstmann}]{BraunAnalysisBreakup2017}
Braun, V., Lemmens, S., Reihs, B., Krag, H., and Horstmann, A., \enquote{Analysis of Breakup Events,} \emph{7th European Conference on Space Debris}, 2017.

\bibitem[{Servadio et~al.(2024)Servadio, Jang, and Linares}]{servadio2024threat}
Servadio, S., Jang, D., and Linares, R., \enquote{Threat Level Estimation From Possible Break-Up Events In LEO,} , 2024.
\newblock \urlprefix\url{https://arxiv.org/abs/2407.18197}.

\bibitem[{Weeden(2010)}]{SWFiridiumcosmos}
Weeden, B., \enquote{2009 Iridium-Cosmos Collision Fact Sheet,} , 2010.
\newblock \urlprefix\url{https://swfound.org/media/6575/swf_iridium_cosmos_collision_fact_sheet_updated_2012.pdf}.

\bibitem[{{Pultarova}(2023)}]{starlinkManeuver}
{Pultarova}, T., \enquote{{SpaceX Starlink satellites} had to make 25,000 collision-avoidance maneuvers in just 6 months — and it will only get worse,} \url{https://www.space.com/starlink-satellite-conjunction-increase-threatens-space-sustainability}, 2023.
\newblock Accessed: 2023-07-07.

\bibitem[{Moomey et~al.(2023)Moomey, Falcon, and Khan}]{MOOMEY2023217}
Moomey, L. C.~D., Falcon, R., and Khan, A., \enquote{Trending and analysis of payload vs. All low earth conjunction data messages below 1000 km, from 2016 through 2021,} \emph{Journal of Space Safety Engineering}, Vol.~10, No.~2, 2023, pp. 217--230.
\newblock \doi{10.1016/j.jsse.2023.03.006}.

\bibitem[{Liou et~al.(2013)Liou, Anilkumar, Virgili, Hanada, Krag, Lewis, Raj, Rao, Rossi, and Sharma}]{iadccomparison}
Liou, J., Anilkumar, A., Virgili, B.~B., Hanada, T., Krag, H., Lewis, H., Raj, M., Rao, M., Rossi, A., and Sharma, R., \enquote{Stability of the Future LEO Environment - an IADC Comparison Study,} \emph{Proc. 6th European Conference on Space Debris}, 2013.
\newblock \doi{10.13140/2.1.3595.6487}.

\bibitem[{Reuther et~al.(2018)Reuther, Kepner, Byun, Samsi, Arcand, Bestor, Bergeron, Gadepally, Houle, Hubbell, Jones, Klein, Milechin, Mullen, Prout, Rosa, Yee, and Michaleas}]{Reuther2018InteractiveAnalysis}
Reuther, A., Kepner, J., Byun, C., Samsi, S., Arcand, W., Bestor, D., Bergeron, B., Gadepally, V., Houle, M., Hubbell, M., Jones, M., Klein, A., Milechin, L., Mullen, J., Prout, A., Rosa, A., Yee, C., and Michaleas, P., \enquote{{Interactive Supercomputing on 40,000 Cores for Machine Learning and Data Analysis},} \emph{2018 IEEE High Performance Extreme Computing Conference, HPEC 2018}, 2018.
\newblock \doi{10.1109/HPEC.2018.8547629}.

\bibitem[{Liou(2012)}]{LiouLEGENDlecture2012}
Liou, J.-C., \enquote{Orbital Debris Modeling and the Future Orbital Debris Environment,} Orbital Debris Lecture (ASEN 6519), 2012.
\newblock \urlprefix\url{https://ntrs.nasa.gov/api/citations/20120015539/downloads/20120015539.pdf}.

\bibitem[{Liou and Johnson(2005)}]{LEGENDpmd2005}
Liou, J.-C., and Johnson, N., \enquote{A LEO satellite postmission disposal study using LEGEND,} \emph{Acta Astronautica}, Vol.~57, 2005, pp. 324--329.
\newblock \doi{10.1016/j.actaastro.2005.03.002}.

\bibitem[{Liou et~al.(2010)Liou, Johnson, and Hill}]{Liou2010ControllingRemoval}
Liou, J.~C., Johnson, N.~L., and Hill, N.~M., \enquote{{Controlling the growth of future LEO debris populations with active debris removal},} \emph{Acta Astronautica}, Vol.~66, No. 5-6, 2010, pp. 648--653.
\newblock \doi{10.1016/J.ACTAASTRO.2009.08.005}.

\bibitem[{Liou and Krisko(2013)}]{LEGENDpmd2013}
Liou, J.-C., and Krisko, P., \enquote{{An Update on the Effectiveness of Postmission Disposal in LEO},} \emph{64th International Astronautical Congress (IAC)}, 2013.

\bibitem[{Lewis et~al.(2009{\natexlab{b}})Lewis, Swinerd, Newland, and Saunders}]{LewisActiveDAMAGE}
Lewis, H.~G., Swinerd, G.~G., Newland, R.~J., and Saunders, A., \enquote{{Active Removal Study for On-Orbit Debris Using DAMAGE},} \emph{5th European Conference on Space Debris}, 2009{\natexlab{b}}.

\bibitem[{{Lewis} and {Marsh}(2021)}]{Lewis2021DeepTime}
{Lewis}, H.~G., and {Marsh}, N., \enquote{{Deep Time Analysis of Space Debris and Space Sustainability},} \emph{8th European Conference on Space Debris}, 2021, 153.

\bibitem[{White and Lewis(2014{\natexlab{b}})}]{White2014TheRemoval}
White, A.~E., and Lewis, H.~G., \enquote{{The many futures of active debris removal},} \emph{Acta Astronautica}, Vol.~95, No.~1, 2014{\natexlab{b}}, pp. 189--197.
\newblock \doi{10.1016/J.ACTAASTRO.2013.11.009}.

\bibitem[{Virgili(2016)}]{Virgili2016}
Virgili, B.~B., \enquote{DELTA (Debris Environment Long-Term Analysis),} \emph{6th International Conference on Astrodynamics Tools and Techniques (ICATT)}, 2016.

\bibitem[{{ESA Space Debris Office}(2023)}]{ESAspaceEnvReport2023}
{ESA Space Debris Office}, \enquote{{ESA} Annual Space Environment Report,} , 9 2023.
\newblock \urlprefix\url{https://www.esa.int/Space_Safety/ESA_s_Space_Environment_Report_2023}.

\bibitem[{{Letizia} et~al.(2023){Letizia}, {Bella}, {Weber}, {Bastida Virgili}, {Lemmens}, and {Soares}}]{2023Letizia}
{Letizia}, F., {Bella}, D., {Weber}, D., {Bastida Virgili}, B., {Lemmens}, S., and {Soares}, T., \enquote{{Long-term Environment Simulations and Risk Characterization in Support of a Zero Debris Policy},} \emph{2nd International Orbital Debris Conference}, LPI Contributions, Vol. 2852, 2023, p. 6029.

\bibitem[{{Bastida Virgili} et~al.(2016){Bastida Virgili}, Dolado, Lewis, Radtke, Krag, Revelin, Cazaux, Colombo, Crowther, and Metz}]{iadc2016riskToSpaceSustain}
{Bastida Virgili}, B., Dolado, J., Lewis, H., Radtke, J., Krag, H., Revelin, B., Cazaux, C., Colombo, C., Crowther, R., and Metz, M., \enquote{Risk to space sustainability from large constellations of satellites,} \emph{Acta Astronautica}, Vol. 126, 2016, pp. 154--162.
\newblock \doi{10.1016/j.actaastro.2016.03.034}.

\bibitem[{D’Ambrosio et~al.(2022)D’Ambrosio, Servadio, Siew, Jang, Lifson, and Linares}]{MOCAT-AAS-2022}
D’Ambrosio, A., Servadio, S., Siew, P.~M., Jang, D., Lifson, M., and Linares, R., \enquote{Analysis Of The Leo Orbital Capacity Via Probabilistic Evolutionary Model,} \emph{AAS/AIAA Astrodynamics Specialist Conference}, 2022.

\bibitem[{Jang et~al.(2022)Jang, D’Ambrosio, Lifson, Pasiecznik, and Linares}]{MOCAT-bin-AMOS2022}
Jang, D., D’Ambrosio, A., Lifson, M., Pasiecznik, C., and Linares, R., \enquote{Stability of the LEO Environment as a Dynamical System,} \emph{Advanced Maui Optical and Space Surveillance Technologies (AMOS) Conference}, 2022.

\bibitem[{McDowell(2023)}]{mcdowell2023jonathan}
McDowell, J., \enquote{Jonathan's Space Report,} , 2023.
\newblock \urlprefix\url{https://planet4589.org/space/con/conlist.html}.

\bibitem[{Henry(2019)}]{henry2019spacex}
Henry, C., \enquote{SpaceX submits paperwork for 30,000 more Starlink satellites,} \emph{Space News}, Vol.~15, 2019, p. 2019.

\bibitem[{Diaz et~al.(2023)Diaz, Mesalles~Ripoll, Duncan, Lindsay, Harris, and Lewis}]{diaz2023data}
Diaz, P., Mesalles~Ripoll, P., Duncan, M., Lindsay, M., Harris, T., and Lewis, H.~G., \enquote{Data-Driven Lifetime Risk Assessment and Mitigation Planning for Large-Scale Satellite Constellations,} \emph{The Journal of the Astronautical Sciences}, Vol.~70, No.~4, 2023, p.~21.
\newblock \doi{https://doi.org/10.1007/s40295-023-00384-w}.

\end{thebibliography}
